\documentclass[preprint,11pt]{elsarticle}
\usepackage[a4paper,left=1.8cm,right=1.8cm,top=2.2cm,bottom=2.3cm]{geometry}

\usepackage{amsmath,amssymb,amsfonts,bm}
\usepackage{graphicx}
\usepackage{float}      
\usepackage{subcaption}
\usepackage{booktabs}
\usepackage{siunitx}
\usepackage[hidelinks]{hyperref}

\usepackage{ragged2e}
\usepackage{tabularx}

\usepackage{soul}
\usepackage{xcolor}

\usepackage{tabularx}

\graphicspath{{images/}}   
\usepackage{soul}

\begin{document}

\begin{frontmatter}
\title{A Differentiable, GPU-Accelerated Finite-Element
Formulation in JAX for Large-Deformation Shells with Contact:
Application to Patient-Specific Aortic Valve Simulation}

\author[inst1,inst2,inst3]{Mohsen Nakhaei\corref{cor1}}
\ead{nakhaeim@chop.edu}
\author[inst2]{Alison M. Pouch}
\author[inst3]{Matthew A. Jolley}
\author[inst4,inst1,inst5]{Wensi Wu\corref{cor1}}
\ead{wensiwu@engineering.upenn.edu}
\cortext[cor1]{Corresponding authors}

\affiliation[inst1]{organization={Department of Pediatrics, Division of Cardiology,
Children's Hospital of Philadelphia}, city={Philadelphia}, state={PA}, country={USA}}
\affiliation[inst2]{organization={Department of Radiology and Bioengineering,
University of Pennsylvania}, city={Philadelphia}, state={PA}, country={USA}}
\affiliation[inst3]{organization={Department of Anesthesiology and Critical Care Medicine,
Children's Hospital of Philadelphia}, city={Philadelphia}, state={PA}, country={USA}}
\affiliation[inst4]{organization={Department of Mechanical Engineering and Applied Mechanics,
University of Pennsylvania}, city={Philadelphia}, state={PA}, country={USA}}
\affiliation[inst5]{organization={Cardiovascular Institute,
Children's Hospital of Philadelphia}, city={Philadelphia}, state={PA}, country={USA}}

\begin{abstract}
Patient-specific aortic-valve models can help assess how
tissue properties and geometry influence leaflet deformation
and coaptation, with potential to inform intervention planning.
Large deformations, near-incompressibility, and evolving
contact make these simulations computationally demanding
and can challenge nonlinear convergence.
Differentiable, GPU-accelerated methods offer efficient
derivative evaluation and parallel computation, but their
matrix-free application to soft-tissue shells with contact
remains less established.
We developed a forward solver using tatva's energy-based
finite-element framework in JAX. The solver combines finite-strain
extensible-director shells, hyperelastic constitutive models,
and a finite-range repulsive contact energy.
Nonlinear equilibrium is solved using a regularized,
matrix-free Newton--Krylov method with two-level preconditioning.
Within this solver, automatic differentiation (AD) evaluates
tangent--vector products without forming or storing
the primary tangent matrix, while GPU execution
parallelizes element and quadrature-point computations.
Constitutive, shell, contact, and self-contact responses
were assessed against analytical, semi-analytical,
and published benchmarks.
At a contact-free plate equilibrium,
implicit sensitivities of the converged apex displacement
with respect to
constitutive parameters, thickness, and pressure agreed
with finite differences from independent nonlinear simulations.
Simulations of three patient-specific aortic-valve models
(two tricuspid and one bicuspid) achieved final relative
equilibrium residuals below \(10^{-4}\).
Median GPU runtimes were approximately 15--25 minutes,
with 5.3--6.5-fold speedups relative to corresponding
CPU runs on the tested hardware.
These results provide component-level verification and
demonstrate GPU-accelerated forward valve simulation in
the tested cases, supporting future
gradient-based inverse studies.

\end{abstract}

\begin{keyword}
Differentiable finite elements \sep
Automatic differentiation \sep
Matrix-free Newton--Krylov \sep
Finite-strain shells \sep
Contact mechanics  \sep
Patient-specific aortic valve 
\end{keyword}

\end{frontmatter}


\section{Introduction}\label{sec:intro}

Aortic valve (AV) disease is among the most common forms
of valvular heart disease and a frequent reason for
surgical or transcatheter intervention
\citep{Otto2021ACCVHD,dArcy2021Epidemiology}.
Patient-specific computational models offer a means
of examining how leaflet structure and deformation influence
valve function.
These models provide estimates of leaflet stresses and
strains that can inform biomechanical assessment of disease
and potential interventions
\citep{labrosse2015subject,rego2022patient}.
Their clinical value depends on simulation accuracy,
computational efficiency, and validation against independent
data \citep{dowling2020firsthuman,
rauseo2026computational}.
During valve closure, thin, nearly incompressible leaflets
undergo large deformations that alter their geometry and
tangent stiffness, and contact forces develop as the
leaflets approach one another
\citep{marom2012fluid,kamensky2017contact}.
Solving the resulting nonlinear equilibrium equations
requires repeated evaluation of forces and their derivatives.
A differentiable finite-element formulation supports these
calculations in the forward solve and provides a foundation
for future gradient-based inverse studies.
GPU execution can accelerate element and quadrature-point
computations through parallel processing, potentially making
repeated patient-specific simulations more practical for
clinical assessment and intervention planning.

Automatic differentiation (AD) provides these derivative
calculations by applying the chain rule to implemented
energy and residual functions
\citep{vigliotti2021ad,pundir2026versatile}.
Together with GPU computing, it supports both the training
of neural-network finite-element (NNFE) models and
Newton-type solution of finite-element equilibrium equations.
For example, \citet{meyer2025nnfe} developed an NNFE model
of three-dimensional solid valve leaflets with multi-body
contact.
After energy-based training, the network predicts
displacements over the pressure range considered.
In Newton-type finite-element solvers, AD differentiates
the residual with respect to the finite-element unknowns
to evaluate tangent operators or tangent--vector products,
reducing the need for manually derived expressions
\citep{vigliotti2021ad}.
JAX-FEM and torch-fem implement differentiable finite-element
formulations for structural analysis and gradient-based
optimization \citep{xue2023jaxfem,meyer2024torchfem}.
TensorMesh uses tensorized finite-element assembly and
sparse linear solvers for GPU-based computation.
The associated TensorGalerkin framework supports AD through
assembly and adjoint differentiation of sparse linear solves
\citep{wen2026tensorgalerkin}.
In biomechanics, PyTorch-FEA applies AD to forward inflation
and inverse identification of patient-specific material
parameters in the human aorta
\citep{liang2023pytorchfea}.

Within the JAX ecosystem, tatva uses an energy-centric
finite-element formulation in which AD computes
residuals and tangent operators from the global energy
\citep{pundir2026versatile}.
This structure allows us to express shell deformation
and contact as separate energy terms and combine them before
differentiation.
When the material or contact energy expressions are
modified, AD computes the corresponding residual and
tangent contributions without manually deriving new
expressions.
Tatva provides reusable operators for field interpolation, gradient evaluation at quadrature
points, and numerical integration, along with utilities
for imposing prescribed boundary values.
For equilibrium solves, tatva evaluates tangent--vector
products without forming the full tangent matrix and provides
graph-coloring-based sparse reconstruction when an explicit
sparse tangent is needed.
JAX supplies the underlying automatic differentiation,
vectorization, and execution on CPUs and GPUs
\citep{bradbury2018jax}.
Together, tatva's finite-element operators and JAX's
computational capabilities provide a suitable foundation for
our valve application.

However, applying these computational capabilities to valve
mechanics also requires suitable shell discretization,
treatment of evolving contact,
and an effective nonlinear solution procedure.
In shell discretizations, constraints on shear and
through-thickness deformation can introduce artificial
stiffness \citep{leebathe2004mitc3,buchter1994three}.
Large stiffness differences can make the linearized
equilibrium equations difficult to solve, making effective
preconditioning important for iterative methods
\citep{knoll2004jfnk}. Evolving contact also presents challenges for differentiability
because the interacting surface locations and contact status
depend on deformation.
Closest-point projection requires careful treatment of surface
geometry and transitions between face, edge, and vertex
projections
\citep{konyukhov2008solvability,ankit2027geometrically}.
Changes in active contact constraints or the selected closest
facet can introduce nonsmooth behavior, motivating specialized
or regularized contact formulations for sensitivity analysis
\citep{wriggers2006contact,geilinger2020add}.
Addressing these requirements would support repeated forward
simulations and provide a foundation for future gradient-based
inverse studies using implicit differentiation
\citep{blondel2022implicit}.

In the present study, we develop shell and contact
implementations for large-deformation soft-tissue mechanics within
tatva's energy-based finite-element framework in JAX.
Our implementation combines finite-strain extensible-director
shells, hyperelastic constitutive models, and a finite-range
quartic repulsive contact energy.
The three-node shell incorporates mixed interpolation of tensorial
components (MITC3) to reduce transverse-shear locking
\citep{leebathe2004mitc3} and an enhanced-assumed-strain
(EAS) mode to enrich the transverse-normal strain field
\citep{buchter1994three}.
Nonlinear equilibrium is solved using a regularized,
matrix-free Newton--Krylov method with two-level preconditioning.
We first evaluate constitutive response, shell deformation,
and contact behavior against analytical and semi-analytical
reference solutions and published benchmarks, including a
self-contact ring test.
We subsequently verify implicit derivatives of the converged
apex displacement using a contact-free plate.
Finally, we apply the formulation to three patient-specific
aortic valves reconstructed from image segmentations and examine
mesh and numerical sensitivity. CPU and GPU performance are
compared for the plate and valve simulations.

Section~\ref{sec:methods} presents the mechanical formulations
and nonlinear solution strategy.
Section~\ref{sec:results} presents component verification,
CPU--GPU performance comparisons,
and patient-specific valve-closure results.
Section~\ref{sec:discussion} discusses the main
findings and limitations.

\section{Methods}\label{sec:methods}

\subsection{Governing equations and constitutive models}
\label{sec:gov}

We model deformable bodies as hyperelastic continua
undergoing finite deformation. Let $\Omega_0\subset\mathbb{R}^3$ denote
the reference configuration, $\bm X\in\Omega_0$ the material
coordinates, and
$\bm\varphi(\bm X)=\bm X+\bm u(\bm X)$ the deformation map, where
$\bm u$ is the displacement field.
The deformation gradient, its
Jacobian, and the right Cauchy--Green tensor are
\[
\bm F=\bm I+\nabla_0\bm u,\qquad
J=\det\bm F>0,\qquad
\bm C=\bm F^\top\bm F,
\]
where $\nabla_0$ denotes the gradient with respect
to the reference coordinates.
The first invariant of $\bm C$ is
$I_1=\operatorname{tr}\bm C
=\lambda_1^2+\lambda_2^2+\lambda_3^2$.
It is equal to the sum of the squared principal stretches and
provides the isotropic deformation measure used in the strain-energy
functions below.

For fixed external forces, equilibrium requires
the first variation of the total potential energy to vanish,
consistent with the principle of virtual work.
Writing the internal, contact, and external
contributions separately,
\begin{equation}
\begin{aligned}
\Pi(\bm u)
&=
\Pi_{\mathrm{int}}(\bm u)
+\Pi_{\mathrm c}(\bm u)
-\Pi_{\mathrm{ext}}(\bm u),\\
\Pi_{\mathrm{int}}
&=
\Pi_{\mathrm{mat}}+\Pi_{\mathrm d},
\qquad
\Pi_{\mathrm{mat}}
=
\int_{\Omega_0}\psi(\bm F)\,\mathrm dV ,
\end{aligned}
\label{eq:energy}
\end{equation}
the equilibrium configuration satisfies $\delta\Pi=0$
for all virtual displacements that vanish
where displacements are prescribed.
Here, $\Pi_{\mathrm{mat}}$ is the stored strain energy
of the material, and $\Pi_{\mathrm d}$ is the
shell drilling-stabilization energy
defined in Section~\ref{sec:shell}.
This numerical penalty stabilizes rotation
about the shell's local normal, which otherwise introduces
a zero-energy mode \citep{hughes1989drilling}.
Together, these terms form the internal energy
$\Pi_{\mathrm{int}}$.
The term $\Pi_{\mathrm c}$ is the piecewise-smooth
contact energy described in
Section~\ref{sec:contact}.
The term $\Pi_{\mathrm{ext}}$ represents the work of
the applied loads. Pressure acts normal to the
deformed surface and changes direction as
the surface deforms.
In the pressure-loaded shell simulations,
the loaded-face geometry and pressure force vector are held
fixed within each load step and updated between load steps
and during final equilibrium refinement, as described in
Section~\ref{sec:impl}.
Equation~\eqref{eq:energy} therefore defines the potential
used for fixed-force evaluations; a globally defined potential
for the geometry-dependent pressure load is not assumed.

After finite-element discretization, the potential depends on the
primary nodal variables $\bm U$.
For shells, these include displacements and
rotations, and the potential also depends on
the nodal internal variables $\bm w$
that control the thickness stretch and the element-local
enhanced-strain variables $\bm\alpha$.
These shell variables are defined
in Section~\ref{sec:shell}:
\begin{equation}
  \widetilde\Pi_h(\bm U,\bm w,\bm\alpha)
  =
  \sum_e\sum_q
  \psi_h\!\left(
    \bm U,\bm w,\bm\alpha;\xi_q
  \right)
  \omega_q\det\bm J_e(\xi_q)
  +\Pi_{\mathrm d}
  +\Pi_{\mathrm c}(\bm U)
  -\Pi_{\mathrm{ext}}(\bm U),
  \label{eq:discrete-energy}
\end{equation}
where $\psi_h$ is the strain-energy density,
$\omega_q$ and $\xi_q$ are the quadrature weights
and locations, and $\bm J_e$ is the Jacobian
of the mapping from element coordinates to the reference
configuration.
The notation $\omega_q$ is used for the
quadrature weights to distinguish them from the nodal thickness-stretch
variables $\bm w$.

During each derivative evaluation, $\bm w$, $\bm\alpha$, and the
pressure force vector are held fixed.
For each contact sampling point,
automatic differentiation uses the distance to the target facet that is
closest at the current configuration.
The closest facet may change
between evaluations, making the contact contribution piecewise smooth.
The equilibrium residual $\bm r$ and tangent--vector product $\bm K\bm v$
for an arbitrary direction $\bm v$ are,
\[
\bm r
=
\frac{\partial\widetilde\Pi_h}{\partial\bm U},
\qquad
\bm K\bm v
=
\frac{\partial^2\widetilde\Pi_h}{\partial\bm U^2}\bm v.
\]
Both quantities are evaluated by automatic differentiation, without a
manually derived tangent or assembly of the global
tangent matrix $\bm K$.
The shell discretization is described in Section~\ref{sec:shell}, and
the differentiation and solution procedures are presented in
Section~\ref{sec:impl}.
Because the contact energy is included
in the same scalar functional,
its contributions to $\bm r$ and $\bm K\bm v$ are obtained through the
same automatic-differentiation procedure, without separate
contact-element assembly.
During the final equilibrium refinement, the
coupling between $\bm U$ and the internal shell variables
$\bm w$ and $\bm\alpha$ is also accounted for, as described in
Section~\ref{sec:impl}.
The implementation builds on tatva's energy-based
finite-element approach and its Element base class,
using JAX to evaluate the problem-specific energy gradients
and tangent--vector products \citep{pundir2026versatile}.

\par\medskip
\noindent\textit{Constitutive models:}\quad Three constitutive models are implemented:
a compressible neo-Hookean law for isotropic elasticity,
an isotropic Lee--Sacks law for nonlinear
strain stiffening, and a neo-Hookean ground matrix with
Holzapfel--Gasser--Ogden (HGO) fiber reinforcement
for fiber-induced anisotropy.
Each model penalizes volume changes through a term proportional
to $(\ln J)^2$.
Near-incompressibility is therefore imposed
approximately through a penalty rather than exactly as a
kinematic constraint.

The compressible neo-Hookean strain-energy
density is,
\begin{equation}
  \psi_{\mathrm{NH}}
  =
  \tfrac12\mu\,(I_1-3-2\ln J)
  +
  \tfrac12\lambda(\ln J)^2 ,
  \label{eq:nh}
\end{equation}
where $\mu$ and $\lambda$ are the Lam\'e parameters.
The corresponding infinitesimal bulk modulus is
$K_{\mathrm{bulk}}=\lambda+2\mu/3$.

The isotropic Lee--Sacks law
\citep{lee2014leesacks,wu2018leesacks}
augments a neo-Hookean ground matrix with an exponential
term that phenomenologically represents the progressive strain
stiffening associated with collagen recruitment:
\begin{equation}
  \psi_{\mathrm{LS}}
  =
  \tfrac12 c_0\,(I_1-3-2\ln J)
  +
  \tfrac12 c_1
  \left[
      \exp\!\big(c_2(I_1-3)^2\big)-1
  \right]
  +
  \tfrac12K(\ln J)^2 .
  \label{eq:leesacks}
\end{equation}
For the patient-specific valve simulations,
the material parameters are
$(c_0,c_1,K)=(67,13,5000)$ $\mathrm{kPa}$,
with dimensionless $c_2=38$.
These values were adopted from prior
heart-valve studies
\citep{wu2018leesacks,nakhaei2026registration}.

To represent anisotropy, each collagen fiber
family $i$, with reference orientation $\bm a_{0i}$,
contributes an exponential HGO reinforcement term
\citep{holzapfelGO2000}:
\begin{equation}
  \psi_{\mathrm{aniso}}
  =
  \sum_i\frac{k_1}{2k_2}
  \left[
    \exp\!\left(
      k_2\left\langle I_{4i}-1\right\rangle_+^2
    \right)-1
  \right],
  \qquad
  I_{4i}
  =
  \bm a_{0i}\cdot\bm C\,\bm a_{0i},
  \label{eq:hgo}
\end{equation}
where $\langle x\rangle_+=\max(x,0)$.
Here, $\bm a_{0i}$ is a unit vector,
$k_1$ is a fiber stiffness parameter with units of stress,
and $k_2$ is a dimensionless stiffening parameter.
The complete HGO strain-energy density is,
\begin{equation}
  \psi_{\mathrm{HGO}}
  =
  \psi_{\mathrm{NH}}+\psi_{\mathrm{aniso}}.
  \label{eq:hgo-total}
\end{equation}
The Macaulay bracket activates each fiber family only when
$I_{4i}>1$, corresponding to extension beyond its reference length.
Thus, the fibers carry tensile load but do not resist compression.
The resulting anisotropic energy is continuously
differentiable but not twice continuously differentiable at
$I_{4i}=1$.
The full fiber invariant $I_{4i}$ is used rather than its
isochoric projection~\citep{nolan2014robust}.


\subsection{Shell formulation}\label{sec:shell}

\noindent\textit{Element and kinematics:}\quad We use three-node triangular shell
elements \citep{leebathe2004mitc3}.
The element adopts seven-parameter extensible-director shell
kinematics \citep{buchter1994three,bischoff1997shear}.
The current position of a point in the shell is described by
\begin{equation}
\bm x(\xi,\zeta)
=
\bm x_m(\xi)
+
\frac{t}{2}\,\zeta\,\widetilde{\bm d}(\xi),
\qquad
\zeta\in[-1,1],
\label{eq:shellkin}
\end{equation}
where $\xi$ denotes the in-plane element
coordinates, \(\bm x_m\) is the deformed midsurface,
\(t\) is the reference thickness, and \(\zeta\) is the
normalized through-thickness coordinate.
The scaled director field is interpolated from the three
element nodes as
\begin{equation}
\widetilde{\bm d}(\xi)
=
\sum_{I=1}^{3}
N_I(\xi)\lambda_{\mathrm t,I}\bm d_I,
\label{eq:scaled-director}
\end{equation}
where \(N_I\) is the linear shape function,
\(\bm d_I\) is the unit director, and
\(\lambda_{\mathrm t,I}\) is the thickness stretch at node \(I\).
Thus, the current nodal thickness is \(t\lambda_{\mathrm t,I}\).

Each node carries seven kinematic parameters:
three translational degrees of freedom \(\bm u_I\),
three rotational parameters \(\bm\theta_I\) that define
the director orientation, and one scalar parameter \(w_I\)
that controls the thickness stretch.
The translation and rotation parameters are primary nodal
unknowns, whereas \(w_I\) is treated as an internal nodal variable.
Finite director rotations are represented
using the exponential map \citep{simovuquoc1986beam},
\begin{equation}
\bm d_I
=
\exp\!\left([\bm\theta_I]_{\times}\right)\bm D_I,
\label{eq:director-update}
\end{equation}
where \(\bm D_I\) is the reference unit director and
\([\bm\theta_I]_{\times}\) is the skew-symmetric matrix
associated with \(\bm\theta_I\).
This rotation preserves the unit length
of \(\bm d_I\). The nodal thickness stretch is defined as
\begin{equation}
\lambda_{\mathrm t,I}=\exp(w_I),
\label{eq:lamexp}
\end{equation}
where \(w_I\) is the nodal logarithmic thickness-stretch parameter.
Shared nodal director and thickness-stretch values
make the interpolated scaled director field continuous across
element boundaries.
The exponential definition guarantees
\(\lambda_{\mathrm t,I}>0\) for every value of \(w_I\),
ensuring positive nodal thickness.

\par\medskip
\noindent\textit{Locking treatments:}\quad The shell formulation uses separate treatments for transverse-shear
locking and the artificial thickness constraint associated
with near-incompressibility. Transverse-shear locking becomes significant in the thin-shell limit,
where standard displacement-based triangular elements develop
artificially high shear stiffness.
We reduce this locking using the MITC3
(Mixed Interpolation of Tensorial Components)
assumed-strain interpolation of \citet{leebathe2004mitc3}.
The covariant transverse-shear strains are sampled at the three
edge midpoints and then re-interpolated over the element.
This interpolation exactly reproduces constant shear fields
on flat elements with parallel directors.

Near-incompressibility requires contraction of the body through its thickness when the
in-plane area increases.
A shell with an inextensible director cannot represent this
thickness change and therefore constrains the transverse-normal strain.
This can make the shell artificially stiff.
The effect is particularly important for
the nearly incompressible parameter values
used with the Lee--Sacks model of Eq.~\eqref{eq:leesacks}.

The extensible director of Eq.~\eqref{eq:shellkin} removes this
restriction by allowing the director length to change.
The through-thickness stretch is represented by the nodal field
\(\lambda_{\mathrm t,I}\).
For a flat element with parallel directors
the corresponding transverse-normal Green--Lagrange strain reduces to
\begin{equation}
E_{33}
=
\frac{1}{2}\left(\lambda_{\mathrm t}^{\,2}-1\right).
\label{eq:e33-flat}
\end{equation}
For general director configurations,
the strain is calculated from the complete interpolated
director field in Eq.~\eqref{eq:scaled-director}.

The strain energy is evaluated using three integration
points in the plane of each triangle and two Gauss--Legendre points
through the thickness.
To represent additional variation through the thickness,
we introduce one enhanced-assumed-strain (EAS) mode
\citep{simo1990eas},
\begin{equation}
E^{\mathrm{cov}}_{\zeta\zeta}
\;\longleftarrow\;
E^{\mathrm{cov}}_{\zeta\zeta}
+
\alpha_e\zeta ,
\label{eq:eas}
\end{equation}
where \(\alpha_e\) is an element-local parameter and \(\zeta\) is the
normalized through-thickness coordinate.  The superscript cov denotes the covariant strain components in the shell's natural coordinate system, which are subsequently mapped to Cartesian components. Because \(\alpha_e\) is local to each element,
it introduces no additional global nodal unknowns.

Together, the nodal thickness stretch and the element-local EAS
parameter allow the transverse-normal stress to relax.
Since \(\alpha_e\) enters the potential only through the strain energy,
stationarity with respect to \(\alpha_e\) gives
\(\partial\widetilde\Pi_h/\partial\alpha_e=0\).
For a flat element, this corresponds to the first-moment condition
\[
\int_{\Omega_e} S_{33}\,\zeta\,\mathrm dV=0 ,
\]
where \(S_{33}\) is the transverse-normal
second Piola--Kirchhoff stress and \(\Omega_e\) is the
reference element volume.
Stationarity with respect to the nodal thickness-stretch parameters
provides the complementary equations governing the mean
thickness response.
Consequently, the transverse-normal stress is relaxed in a
variational sense rather than being set to zero at every
integration point.
This allows the full three-dimensional constitutive model,
including its volumetric penalty, to
be evaluated at the shell integration points.
\par\medskip
\noindent\textit{Drilling stabilization:}\quad Rotation about the shell director is called
drilling rotation.
It does not change the director direction and therefore
has no stiffness in the director-based strain energy,
producing a zero-energy mode.
This mode is regularized using a penalty
\citep{hughes1989drilling}.
For the valve simulations, the penalty coefficient is
\[
k_{\mathrm d}=\gamma c_0t,
\qquad
\gamma=\frac{1}{2},
\]
where \(c_0\) is the ground-matrix shear modulus of the Lee--Sacks
material, \(t\) is the shell thickness, and \(\gamma\) is a
dimensionless penalty factor.
The sum of these drilling penalties over all shell elements
defines the regularization energy \(\Pi_{\mathrm d}\)
appearing in Eq.~\eqref{eq:energy}.
This is a small-rotation stabilization, and
exact invariance of the penalty energy
under finite rigid-body rotations is not assumed.

\par\medskip
\noindent\textit{Internal fields and reduction:}\quad The primary Newton iteration is formulated in terms of the nodal
displacement and rotation variables \(\bm U\), while the
through-thickness fields are updated separately.
The nodal thickness variables \(\bm w\) and element-local
enhanced-strain parameters \(\bm\alpha=(\alpha_e)_e\)
are collected into \(\bm a=(\bm w,\bm\alpha)\).
The discrete potential is written as
\(\widetilde\Pi_h(\bm U,\bm a)\).
These internal fields enter the strain energy but not the
contact or pressure-load terms. For fixed \(\bm U\), \(\bm a^\star(\bm U)\) denotes internal-variable
values satisfying a stationarity condition:
the derivatives of the potential energy with respect to
these variables vanish, so infinitesimal variations produce
no first-order change in energy.
Substituting these values into the potential defines
the reduced potential \(\Pi_h(\bm U)\),
\begin{equation}
\frac{\partial\widetilde\Pi_h}{\partial\bm a}
\bigl(\bm U,\bm a^\star(\bm U)\bigr)
=\bm 0,
\qquad
\Pi_h(\bm U)
=
\widetilde\Pi_h\bigl(\bm U,\bm a^\star(\bm U)\bigr).
\label{eq:condensation}
\end{equation}
The two internal fields are treated differently.
The enhanced-strain parameter \(\alpha_e\) is element-local,
so its stationarity condition is one independent scalar
equation per element.
Each equation is solved locally. During the coupled final refinement,
we use each element's linearized equation to express
the update to \(\alpha_e\) in terms of the displacement,
rotation, and thickness updates.
We then substitute this expression into the remaining
equations, preserving the coupling to \(\alpha_e\). Thus, \(\alpha_e\) does not enter the global system as an
independent unknown.
The thickness field \(\bm w\), however, is shared by
neighboring elements.
Its stationarity condition therefore forms a global sparse
nonlinear system, which is solved using regularized inner
Newton updates.

Eliminating the internal fields has an important consequence
for differentiation.
At an exactly stationary internal state with pressure forces
held fixed, the chain rule applied to
Eq.~\eqref{eq:condensation} gives
\[
\frac{\mathrm d\Pi_h}{\mathrm d\bm U}
=
\frac{\partial\widetilde\Pi_h}{\partial\bm U}
+
\left(\frac{\mathrm d\bm a^\star}{\mathrm d\bm U}\right)^{\!\top}
\frac{\partial\widetilde\Pi_h}{\partial\bm a}
=
\frac{\partial\widetilde\Pi_h}{\partial\bm U},
\]
because
\(\partial\widetilde\Pi_h/\partial\bm a=\bm0\)
at \(\bm a^\star\). At this state, the reduced residual can
be obtained by differentiating \(\widetilde\Pi_h\)
with the internal fields held fixed.
The second derivative also includes the
coupling arising from the dependence of
\(\bm a^\star\) on \(\bm U\). During the load path, a limited number of sequential
updates is used for the internal fields.
Therefore, \(\bm\alpha\) and \(\bm w\) may not yet satisfy
their stationarity conditions exactly during an intermediate
iteration, and the primary residual is evaluated while
holding their current values fixed.
At the final load, a regularized refinement step accounts
for the coupling among the three fields and verifies that
the residuals associated with
\(\bm U\), \(\bm w\), and \(\bm\alpha\)
satisfy their prescribed tolerances, as described in
Section~\ref{sec:impl}.

\par\medskip
\noindent\textit{Reported strain measures:}\quad For interpretation, the tangential Green--Lagrange strain can be written
to first order through the shell thickness as
\begin{equation}
\bm E_{\mathrm{tan}}(\zeta)
\approx
\bm E_m+\frac{t}{2}\zeta\,\bm\kappa,
\qquad
\zeta\in[-1,1],
\label{eq:strainlin}
\end{equation}
where \(\zeta\) is the normalized through-thickness coordinate,
\(\bm E_m\) is the midsurface membrane strain, and \(\bm\kappa\) is
the bending strain coefficient.
The subscript \(\mathrm{tan}\) denotes
the in-plane strain components.
The complete finite-strain tensor, including higher-order
through-thickness terms, is evaluated directly at the shell
integration points and is not truncated using
Eq.~\eqref{eq:strainlin}. Each strain tensor is represented by a scalar strain magnitude,
defined using the norm,
\[
\lVert\bm E\rVert
=
\sqrt{\bm E:\bm E}
=
\left(\sum_{i,j}E_{ij}^{\,2}\right)^{1/2}.
\]
This norm combines all components of the strain tensor into a single
scalar value and is invariant under orthonormal changes of
coordinate frame. The membrane strain \(\lVert\bm E_m\rVert\) characterizes
the in-plane deformation of the shell midsurface.
The strain magnitudes on the upper and lower
shell faces are approximated by
\begin{equation}
\left\lVert\bm E_{\mathrm{tan}}(\zeta=\pm1)\right\rVert
\mathrel{\approx}
\left\lVert
\bm E_m
\pm
\frac{t}{2}\bm\kappa
\right\rVert,
\label{eq:surfstrain}
\end{equation}
which includes both membrane and bending contributions.

Unless stated otherwise, the reported shell strain is obtained
by first taking the quadrature-weighted average of the full
\(3\times3\) Green--Lagrange strain tensor over the six
integration points of each element: two through-thickness
points at each of the three in-plane integration locations.
The norm is then applied to the averaged tensor to obtain
the reported scalar strain magnitude.
For a flat element with a thickness-symmetric mapping,
the equal weights at paired through-thickness
points cancel the linear bending term and the EAS contribution
\(\alpha_e\zeta\) in the tensor average.
On a curved element, however, the mapping from the shell
coordinates to the Cartesian strain tensor varies through
the thickness.
Exact cancellation is therefore not assumed.
Consequently, the reported tensor is a full three-dimensional
element average that includes the transverse-normal
strain and transverse-shear strains,
as well as any remaining higher-order
through-thickness contributions.


\subsection{Contact and self-contact formulation}\label{sec:contact}

Contact is represented by a short-range repulsive
surface energy, inspired by potential-based contact formulations
\citep{sauer2013surfacepotential,kamensky2017contact}.

\par\medskip
\noindent\textit{Distance measure and search:}\quad Each surface triangle contributes one quadrature sample
located at its centroid in the current configuration and
weighted by its reference area \(A_s\).
This one-point rule approximates a surface integral over
the reference surface, whose quadrature
weights remain constant during deformation.
For contact between separate shell surfaces,
each sample uses a list of \(N_{\mathrm c}=6\) candidate
target facets, identified using a KD-tree search based on
the current facet centroids.
The candidate lists are rebuilt after each accepted load step.
Within a load step, the candidate set is fixed, but the facet
that gives the minimum distance is re-evaluated from the current
geometry during every contact-energy evaluation.

For each candidate facet \(T_k\), the closest point \(\bm c_k\)
is computed using barycentric coordinates located
to the triangle face, edges, or vertices \citep{gribanov2018gradient}.
The corresponding regularized point-to-triangle distance is
\begin{equation}
\operatorname{dist}_{\varepsilon}(\bm x_s,T_k)
=
\sqrt{\lVert\bm x_s-\bm c_k\rVert^2+\varepsilon_d^2},
\label{eq:point-triangle-distance}
\end{equation}
where \(\bm x_s\) is the current position of the source-triangle
centroid.
The small value
\(\varepsilon_d=\SI{1e-12}{\milli\metre}\)
prevents an undefined derivative at exactly zero distance
and has a negligible effect away from that point.
The contact distance is the minimum over the candidate facets,
\begin{equation}
d
=
\min_{k=1,\ldots,N_{\mathrm c}}
\operatorname{dist}_{\varepsilon}\bigl(\bm x_s,T_k\bigr).
\label{eq:dist}
\end{equation}
This distance accounts for closest points on
facet edges and vertices, rather than only projections onto
the facet plane.
\par\medskip
\noindent\textit{Contact potential:}\quad  The unsigned distance \(d\) is measured between the shell
midsurfaces. The contact-activation measure is
\begin{equation}
g=\big\langle R_{\mathrm{out}}-d\big\rangle_+,
\qquad
\langle x\rangle_+=\max(x,0),
\label{eq:gap}
\end{equation}
where \(\langle x\rangle_+\) denotes the positive part
of \(x\) and \(R_{\mathrm{out}}\) defines the short interaction range.
The potential is inactive when \(d\geq R_{\mathrm{out}}\)
and becomes active when the midsurface distance $d$ falls below
\(R_{\mathrm{out}}\). The interaction range is a numerical
parameter and does not depend on the local thickness stretch in the
present expression.
The contact contribution to the total potential energy is,
\begin{equation}
\Pi_{\mathrm c}
=
\sum_s \pi_s,
\qquad
\pi_s
=
k_{\mathrm c}A_sg_s^{4},
\label{eq:contact}
\end{equation}
where \(k_{\mathrm c}\) is the
penalty coefficient,
\(A_s\) is the reference area associated with sample \(s\),
and the sum runs over the contact samples defined above.
The quartic positive-part function
\(\langle x\rangle_+^4\) is three times continuously
differentiable, with its value and first three derivatives
vanishing at activation.
Within a fixed closest-point branch, the contact kernel is
\(C^3\), its force is \(C^2\), and its tangent is \(C^1\).
When one candidate facet is uniquely closest and the
closest-point region is unchanged, automatic differentiation
provides the derivative of that branch.
With a fixed candidate list, the distance
remains continuous, but switching the closest facet can
produce a discontinuous gradient.
The complete contact formulation is therefore piecewise smooth.
Because the contact potential is part of the total energy,
its branchwise residual and symmetric Hessian contributions
are obtained automatically using the same differentiation
procedure as for the strain energy.

\par\medskip
\noindent\textit{Parameter continuation and self-contact:}\quad  For the valve simulations, the contact
parameters are ramped linearly with the applied load fraction
during the first half of the loading path and remain at their
final values thereafter.
The penalty coefficient increases from
\(k_{\mathrm c}=\SI{1000}{\kilo\pascal\per\cubic\milli\metre}\)
to \(\SI{8000}{\kilo\pascal\per\cubic\milli\metre}\),
while the interaction range decreases from
\(R_{\mathrm{out}}=\SI{0.8}{\milli\metre}\)
to \(\SI{0.5}{\milli\metre}\).
The wider initial range and lower penalty
coefficient introduce repulsion gradually; reducing the range
and increasing the coefficient then localize and strengthen
the interaction.
This numerical continuation assists nonlinear convergence. The same contact potential was used for self-contact in the
pressurized-ring benchmark.
For each source facet, candidate target facets were restricted
to facets at least five circumferential cells away.
This exclusion prevents spurious repulsion
between nearby facets connected along the ring.
In the patient-specific simulations, candidate target facets
were restricted to the other leaflets; therefore,
leaflet self-contact was not active.



\subsection{Solver and implementation}\label{sec:impl}

The framework is implemented in JAX \citep{bradbury2018jax}, which
provides automatic differentiation, automatic vectorization of the
element and contact kernels through \texttt{vmap}, just-in-time
compilation, and execution on both CPUs and GPUs from a single code
base. The full equilibrium problem comprises the primary and internal
stationarity equations with pressure evaluated on the current geometry.
Writing \(\bm y=(\bm U,\bm w,\bm\alpha)\) and denoting a model or
loading parameter by \(q\), these equations can be collected as
\(\bm R(\bm y,q)=\bm0\). The operators used to obtain corrections
during load advancement and final refinement approximate the Jacobian
of the full equilibrium equations, including the coupling among the
primary and internal variables and the geometry dependence of the
pressure forces. The following description
distinguishes these correction operators from the full equilibrium
Jacobian required for implicit solution sensitivities.

\par\medskip
\noindent\textit{Derivatives and matrix-free operators:}\quad During derivative evaluations for the primary Newton solve,
the internal fields and pressure force vector are held fixed.
Within a smooth contact branch, the discrete potential in
Eq.~\eqref{eq:discrete-energy} is a differentiable scalar function.
Therefore, reverse-mode AD computes the residual
as the gradient of this potential,
\(\bm r=\partial\widetilde\Pi_h/\partial\bm U\),
following the energy-based differentiation approach of the
tatva framework \citep{pundir2026versatile}.
The Hessian--vector product (\(\bm K\bm v\))
is obtained by forward-over-reverse differentiation:
a forward-mode directional derivative is applied to the
reverse-mode residual in the direction \(\bm v\), giving
\begin{equation}
\bm K\bm v
=
\lim_{\epsilon\to0}
\frac{\bm r(\bm U+\epsilon\bm v)-\bm r(\bm U)}
{\epsilon}.
\label{eq:hvp}
\end{equation}
Here, \(\bm K=\partial\bm r/\partial\bm U\)
is the tangent operator with the internal fields and pressure
forces held fixed.
AD evaluates this derivative using the chain rule,
without finite-difference approximations.
With the internal fields and pressure force vector held fixed,
\(\bm K\) is the Hessian of the total potential energy and is symmetric
on each smooth contact branch. This statement concerns the
fixed-field operator, rather than the Jacobian of the complete
geometry-dependent equilibrium problem. The Krylov solve evaluates
the primary tangent action without assembling that operator;
auxiliary matrices are assembled for preconditioning and internal-field
solves, as described below.

\par\medskip
\noindent\textit{Internal fields and the applied operator:}\quad During final equilibrium refinement, the element-local
EAS updates are eliminated from the linearized equations.
The remaining thickness-field coupling is included through
the regularized reduced tangent
\begin{equation}
\widehat{\bm K}_{\mathrm{red}}
=
\bm K_{\bm U\bm U}^{(\alpha)}
-
\bm K_{\bm U\bm w}^{(\alpha)}
\left(
\bm K_{\bm w\bm w}^{(\alpha)}
+\beta\bm I
\right)^{-1}
\bm K_{\bm w\bm U}^{(\alpha)}.
\label{eq:reduced-tangent}
\end{equation}
The subscript \(\mathrm{red}\) denotes reduction to the
primary variables \(\bm U\), and the superscript \((\alpha)\)
indicates that the EAS variables have been eliminated.
The diagonal blocks describe the primary and thickness-field
responses, while the off-diagonal blocks describe their coupling.
The second term accounts for thickness-field changes in the
primary correction. Here, \(\bm I\) is the identity matrix, and \(\beta\) is
\(10^{-2}\) times the mean absolute diagonal entry of
\(\bm K_{\bm w\bm w}^{(\alpha)}\).
The inverse denotes a sparse linear solve and is not formed
explicitly. The shift changes the Newton correction, not the equilibrium
equations. The thickness-field residual is included in the reduced
right-hand side, whereas the EAS residual is checked separately,
as detailed in Table~\ref{tab:internal-update}.
The final correction therefore approximates the system
obtained by eliminating both internal fields exactly.

\par\medskip
\noindent\textit{Load stepping:}\quad The applied pressure is increased incrementally,
with each load step starting from the accepted configuration
of the previous step.
At step \(k\), the pressure force vector is
computed using the loaded-face geometry from
\(\bm U^{(k-1)}\) and held fixed during the nonlinear solve.
The work of the fixed pressure forces is therefore
\begin{equation}
\Pi_{\mathrm{ext}}^{(k)}(\bm U)
=
\bm F_{\mathrm p}^{(k)}\cdot\bm U,
\qquad
\bm F_{\mathrm p}^{(k)}
=
\int_{\Gamma_{\mathrm p}(\bm U^{(k-1)})}
p^{(k)}\,\bm N^{\top}\bm n\,\mathrm da ,
\label{eq:follower}
\end{equation}
where \(p^{(k)}\) is the applied pressure,
\(\bm N\) collects the element shape functions, and
\(\bm n\) and \(\mathrm da\) are the face normal and
area element of the frozen configuration. The load increment is adapted during loading
and reduced when a step is rejected.
At full load, the pressure forces are recalculated from
the updated geometry, and the coupled equilibrium refinement
is repeated.

\par\medskip
\noindent\textit{Newton--Krylov solution:}\quad  Each load step is solved by an inexact Newton--Krylov method
\citep{dembo1982inexact,knoll2004jfnk}, in which the linearized system
\begin{equation}
(\bm K+\mu\bm I)\,\Delta\bm U=-\bm r
\label{eq:newton}
\end{equation}
is solved approximately and matrix-free using the preconditioned
conjugate-gradient (CG) method.
Here, \(\mu>0\) is the regularization parameter
and \(\bm I\) is the identity matrix. CG requires a symmetric positive-definite operator, whereas
the shell tangent may become indefinite because of geometric
nonlinearity and contact.
To stabilize the correction in these cases,
we add the diagonal shift \(\mu\bm I\), known as
Levenberg regularization.
Because the shifted operator can still be indefinite,
CG checks the curvature of the local quadratic energy
along each search direction.
The iteration stops when this quantity is non-positive
or nearly zero \citep{steihaug1983}. The returned correction is tested by a backtracking line search,
which first tests the full update and reduces its step length
if the potential-energy acceptance condition is not satisfied.
If no acceptable step is found, the correction is rejected
and \(\mu\) is increased. Because the Levenberg shift modifies only the Newton correction
and not the equilibrium equations, convergence is assessed
using the original equilibrium residual \(\bm r\),
so the final configuration must satisfy the
original equilibrium equations to the prescribed tolerance.

\par\medskip
\noindent\textit{Preconditioning:}\quad To accelerate Krylov convergence
when the linearized system contains large
stiffness differences, we use a two-level preconditioner combining
a positive diagonal (Jacobi) approximation \(\bm D\)
with a coarse correction:
\begin{equation}
\bm M^{-1}
=
\bm D^{-1}
+
\bm P\,\bm A_{\mathrm c}^{-1}\bm P^{\top}.
\label{eq:prec}
\end{equation}
The diagonal term rescales local stiffness differences,
while the coarse correction represents collective
deformation across groups of nodes.
Here, \(\bm D\) includes internal-energy and approximate
contact contributions, \(\bm P\) maps coarse translations
and rotations to the nodal variables, and
\(\bm A_{\mathrm c}\) is a regularized
coarse approximation constructed from the assembled
internal-energy tangent and approximate contact contributions.
The coarse space uses three translational and three rotational
rigid-body modes for each spatial group of nodes. The diagonal and coarse components both include
the Levenberg shift from Eq.~\eqref{eq:newton}.
The shift-dependent parts are updated at every Newton iteration,
while the assembled internal-energy and contact contributions
are refreshed every fourth iteration.

\par\medskip
\noindent\textit{Load-path and convergence criteria:}\quad The patient-specific simulations are quasi-static and introduce no inertia. A permissive residual tolerance is used to advance the load path,
followed by stricter convergence checks at full load.
Along the load path, the acceptance check after updating
the internal variables is
\[
\lVert\bm r\rVert
<
5\times10^{-2}
\max\!\left(\lVert\overline{\bm F}_{\mathrm p}\rVert,F_0\right),
\]
where \(\overline{\bm F}_{\mathrm p}\) is the nodal force
corresponding to the target pressure, evaluated on the
loaded-surface geometry held fixed during
the load step, and \(F_0=\SI{1}{\milli\newton}\)
prevents a vanishing load scale. At full load, the internal variables and geometry-dependent
pressure force are updated, and the configuration is refined
using the regularized reduced tangent in
Eq.~\eqref{eq:reduced-tangent}.
The final static residual must satisfy
\[
\lVert\bm r\rVert
\le
10^{-3}
\max\!\left(\lVert\bm F_{\mathrm p}\rVert,F_0\right).
\]
The internal-variable and pressure-force convergence checks
must also hold at the same updated state. The pressure-force update
criterion is defined in Section~\ref{sec:verification-audit}.
If these criteria are not met within the prescribed maximum number of refinement iterations,
the simulation is classified as unconverged.
Table~\ref{tab:internal-update} summarizes the updates, with
$\bm r_{\bm U}=\bm r$ denoting the primary equilibrium residual.


\begin{table}[htbp]
\centering
\caption{Algorithm for the matrix-free Newton--Krylov
solution with incremental loading and coupled full-load
equilibrium refinement.}
\label{tab:internal-update}
\small
\begin{tabular}{cp{0.84\linewidth}}
\toprule
Step & Operation and objective \\
\midrule

\multicolumn{2}{p{0.92\linewidth}}{
\textbf{Part 1: Advance to the target load.}
For each load step, compute the pressure force and contact
candidate lists from the previous accepted configuration
and hold them fixed during the step.
Internal equilibrium is approximated using limited Newton updates.

\textbf{Repeat Steps 1--4:}}\\

1 &
\textbf{Update the thickness field.}
At fixed $\bm U$, apply global Newton updates to the
thickness-field equations:
\[
\bm r_{\bm w}
=
\frac{\partial\widetilde\Pi_h}{\partial\bm w}
=
\bm 0.
\]
\\

2 &
\textbf{Update the EAS variables.}
Apply local Newton updates to the internal
equilibrium equation in each element:
\[
r_{\alpha_e}
=
\frac{\partial\widetilde\Pi_h}{\partial\alpha_e}
=
0,
\qquad \forall e.
\]
\\

3 &
\textbf{Compute the primary correction.}
With the internal fields and pressure force fixed,
use preconditioned CG and backtracking:
\[
(\bm K+\mu\bm I)\Delta\bm U=-\bm r_{\bm U},
\qquad
\bm U\leftarrow\bm U+s\Delta\bm U.
\]
Increase $\mu$ and retry if backtracking fails.
\\

4 &
\textbf{Check the load step.}
Refresh the internal fields at the updated $\bm U$
and re-evaluate the equilibrium residual.
\[
\text{\textbf{Until}}\quad
\frac{\lVert\bm r_{\bm U}\rVert}
{\max(\lVert\overline{\bm F}_{\mathrm p}\rVert,F_0)}
<5\times10^{-2}.
\]
Accept the step and advance the load.
If the iteration limit is reached,
reduce the load increment and retry.
\\

\midrule
\multicolumn{2}{p{0.92\linewidth}}{
\textbf{Part 2: Converge the full-load equilibrium.}
At the target pressure, refine the coupled fields
and update the geometry-dependent pressure force.

\textbf{Repeat Steps 5--6:}
}\\

5 &
\textbf{Refine the coupled equilibrium.}
Update the internal fields, eliminate the local EAS coupling,
and use the reduced tangent:
\[
\begin{aligned}
\widehat{\bm K}_{\mathrm{red}}
&=\bm K_{\bm U\bm U}^{(\alpha)} -
\bm K_{\bm U\bm w}^{(\alpha)}
\left(\bm K_{\bm w\bm w}^{(\alpha)}+\beta\bm I\right)^{-1}
\bm K_{\bm w\bm U}^{(\alpha)}.
\end{aligned}
\]
Include the thickness-field residual in the
right-hand side of the correction equations,
apply the Levenberg shift $\mu\bm I$, and update
$\bm U$ and $\bm w$ together using backtracking.
\\

6 &
\textbf{Update pressure forces and check convergence.}
Recalculate
$\bm F_{\mathrm p}=\bm F_{\mathrm p}(\bm U)$
and re-evaluate the original equilibrium residual:
\[
\text{\textbf{Until}}\quad
\frac{\lVert\bm r_{\bm U}\rVert}
{\max(\lVert\bm F_{\mathrm p}\rVert,F_0)}
\le10^{-3},
\]
and the pressure-force and internal-variable
convergence criteria are satisfied
at the same updated state.
\\

\bottomrule
\end{tabular}
\end{table}



\section{Results}\label{sec:results}
This section first presents verification of the constitutive,
contact, and shell components, followed by differentiation
verification and solver-performance comparisons.
The formulation is then applied to three patient-specific
aortic valves, including mesh and numerical-sensitivity checks.

\subsection{Thick-walled cylinder inflation:
constitutive-model verification}
\label{sec:material}

The isotropic neo-Hookean and Lee--Sacks laws and the anisotropic
Holzapfel--Gasser--Ogden (HGO) model were verified using
inflation of a thick-walled cylinder with an
incompressible reference solution. A quarter cylinder with inner radius
$A=\SI{5}{\milli\metre}$ and outer radius
$B=\SI{10}{\milli\metre}$ was modeled under plane-strain
conditions. Symmetry conditions were applied to the radial cut
faces, and the outer wall was traction-free. The inner-wall
loading was controlled by either prescribed radial displacement
or pressure loading. Geometry and material parameters are
summarized in Table~\ref{tab:artery}. The two HGO fiber families lie in the
circumferential--axial plane, oriented at
$\beta=\pm45^\circ$ to the circumferential direction:
\[
\bm a_0^{(1,2)}
=\cos\beta\,\bm e_\theta
\pm\sin\beta\,\bm e_z .
\]
Their fiber invariant under plane strain is
$I_4=\lambda_\theta^2\cos^2\beta+\sin^2\beta$. For the incompressible reference, the current
radius $r$ and principal stretches are
\[
r(R)=\sqrt{R^2+a^2-A^2},\qquad
\lambda_\theta=\frac{r}{R},\qquad
\lambda_r=\frac{R}{r},\qquad
\lambda_z=1 ,
\]
where $R$ is the reference radius and $a$
is the deformed inner radius. The corresponding pressure is
\[
P(a)=\int_a^{b_{\mathrm d}}
\frac{\sigma_{\theta\theta}-\sigma_{rr}}{r}\,\mathrm dr ,
\]
where $b_{\mathrm d}=r(B)$ is the deformed outer radius.
This one-dimensional equilibrium integral
provides a reference independent of the finite-element
discretization. It was evaluated using $600$ uniformly spaced
points; increasing this number to $2400$ changed the pressure
by at most $0.00062\%$.

\textit{In the displacement-controlled tests}, the inner radius $a$
was prescribed and the resulting pressure was compared with
$P(a)$. The mean relative pressure errors were $0.05\%$,
$0.15\%$, and $0.06\%$ for the neo-Hookean, Lee--Sacks,
and HGO models, respectively.

\begin{figure}[H]
\centering
\includegraphics[width=\linewidth]{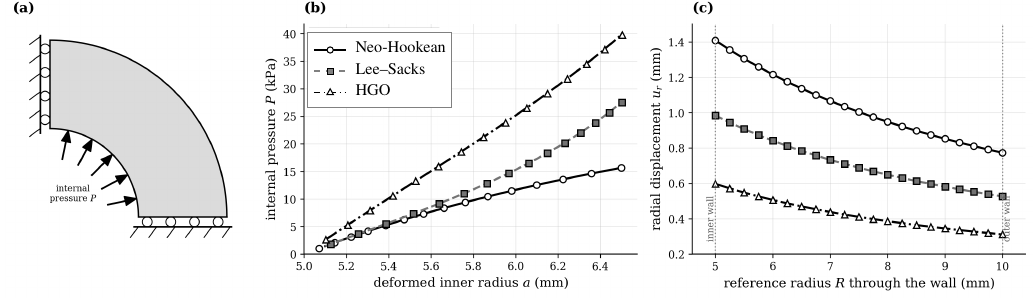}
\caption{Finite-element results are compared with the
incompressible reference solutions for thick-walled cylinder inflation.
(a)~Boundary conditions; (b)~internal pressure versus deformed
inner radius; and (c)~radial displacement through the wall.
Lines show reference solutions and markers show finite-element results.}
\label{fig:artery-pd}
\end{figure}

\textit{In the pressure-controlled tests}, the computed inner radius
differed from the reference by less than $0.05\%$ over the
tested range (Fig.~\ref{fig:artery-pd}b).
At \SI{15}{\kilo\pascal}, the radial displacement
profile agreed with the reference to within $0.01\%$, and the
HGO fiber invariant agreed to within $0.04\%$.
These results verify the three constitutive implementations
with the near-incompressible plane-strain discretization.

\begin{table}[H]
\centering
\caption{Geometry and material parameters for the thick-walled
cylinder verification.}
\label{tab:artery}
\small
\begin{tabular}{ll}
\toprule
Model & Parameters \\
\midrule
Geometry and kinematics
& $A=\SI{5}{\milli\metre}$, $B=\SI{10}{\milli\metre}$,
  $\lambda_z=1$ \\

neo-Hookean
& $\mu=\SI{50}{\kilo\pascal}$,
  $\lambda=\SI{5e4}{\kilo\pascal}$ \\

Lee--Sacks
& $c_0=\SI{50}{\kilo\pascal}$,
  $c_1=\SI{25}{\kilo\pascal}$,
  $c_2=5$,
  $K=\SI{5e4}{\kilo\pascal}$ \\

HGO
& $\mu=\SI{50}{\kilo\pascal}$,
  $\lambda=\SI{5e4}{\kilo\pascal}$,
  $k_1=\SI{75}{\kilo\pascal}$,
  $k_2=4$,
  $\beta=\ang{45}$ \\
\bottomrule
\end{tabular}
\end{table}

\subsection{Verification of the contact formulations}
\label{sec:traction}

Three benchmarks examined different aspects of contact.
Prescribed-motion analyses verified the quartic contact law and its
numerical integration. A contact patch test examined force transmission
using separate quadratic-penalty and augmented-Lagrangian formulations
\citep{simo1992augmented,zimmerman2018surfacecontact}. A pressurized-ring
benchmark then assessed coupled shell deformation and self-contact using
the same quartic formulation as the valve simulations, with equilibrium
obtained through direct energy minimization.

\subsubsection{Quartic contact formulation}

\begin{figure}[H]
\centering
\includegraphics[width=\linewidth]{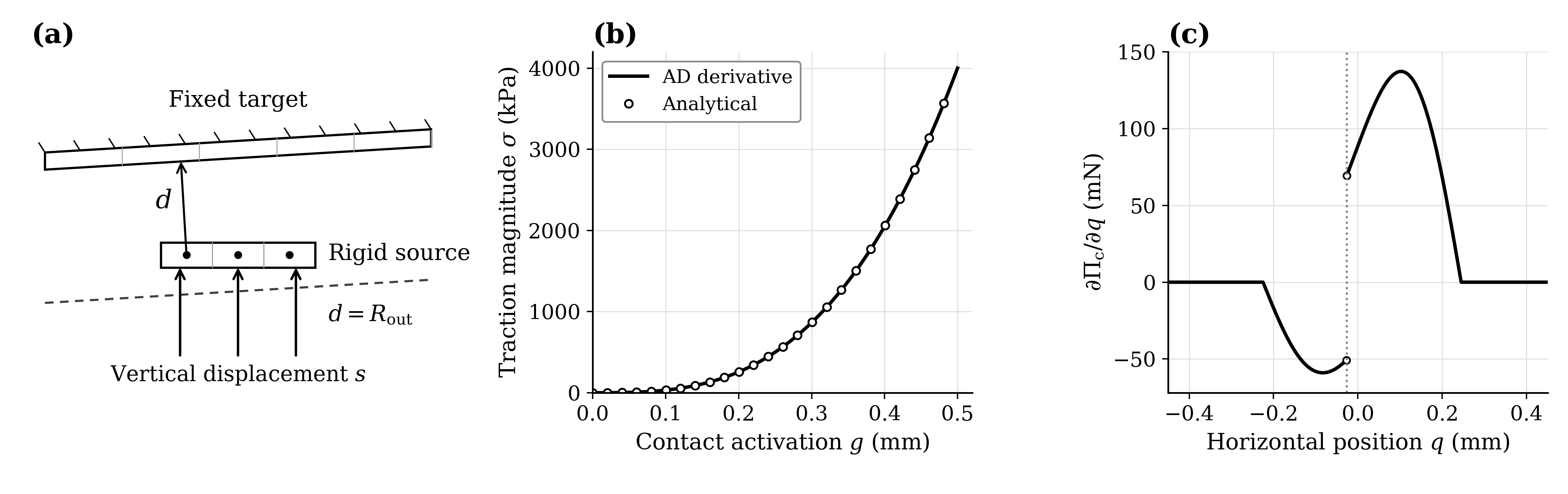}
\caption{Verification of the quartic contact formulation.
(a)~Schematic of the contact configuration;
(b)~analytical and AD traction magnitudes; and
(c)~contact-energy gradient
$\partial\Pi_{\mathrm c}/\partial q$ versus horizontal
source-point position $q$ in the separate geometric test.
The dotted line marks the change
in the closest target triangle.}
\label{fig:traction}
\end{figure}

The quartic contact law was examined through traction verification,
surface-integration accuracy, and
the contact response to changes in the closest
target triangle.
A rigid $4\times4$~mm source surface
was centered below a fixed $10\times10$~mm target
surface (Fig.~\ref{fig:traction}a).
The target had a slope of $0.02$ in the $x$ direction,
and the initial central vertical separation was
$d_0=\SI{0.30}{\milli\metre}$.
The source underwent a prescribed displacement $s$
in the $z$ direction.
Both surfaces were triangulated.
For each source-triangle centroid, $d$ was the shortest
distance to the closest of six candidate target triangles.
Because the target was tilted, this local distance varied
across the source surface. Differentiating the quartic energy in
Eq.~\eqref{eq:contact} gives the analytical traction magnitude
during active contact,
\[
\sigma=4k_{\mathrm c}g^3,
\qquad
g=\langle R_{\mathrm{out}}-d\rangle_+.
\]
The contact parameters were
$k_{\mathrm c}=\SI{8000}{\kilo\pascal\per\cubic\milli\metre}$
and $R_{\mathrm{out}}=\SI{0.5}{\milli\metre}$.
The AD derivative of the quartic law agreed with this
analytical expression, with a maximum relative error of
$2.2\times10^{-16}$ (Fig.~\ref{fig:traction}b).

To verify the numerical integration of contact energy and
force over the source surface, the source mesh was refined
while keeping the tilted target geometry fixed.
At each refinement step, the source facet size was reduced
to half its previous value.
The computed contact energy and total contact force were
compared with analytical surface integrals.
Both relative errors decreased by approximately a factor
of four at each refinement step.
Thus, the centroid-based integration approached the
analytical energy and force with second-order accuracy
for this geometry.

A separate geometric test used a single moving source point
and two fixed, non-deforming target triangles.
The triangles represented two separated regions of the
target surface, with a gap between them.
The source point moved horizontally above them along
$\bm x_{\mathrm p}(q)=(q,0,0)$,
where $q$ denotes its horizontal position.
The test examined whether the contact force remained
continuous when the nearest target changed from one
triangle to the other.
The contact distance was
\[
d(q)=\min\bigl(d_1(q),d_2(q)\bigr),
\]
where $d_1(q)$ and $d_2(q)$ are the shortest
distances from the source point to the first and second
target triangles, respectively.
Using a reference area of
$\SI{1}{\square\milli\metre}$ for the source point,
Fig.~\ref{fig:traction}c plots
$\partial\Pi_{\mathrm c}/\partial q$, which measures
how the contact energy changes as the source point moves.
The horizontal contact force on the point is the negative
of this derivative:
\[
F_x=-\frac{\partial\Pi_{\mathrm c}}{\partial q}.
\]
At $q\approx\SI{-0.0266}{\milli\metre}$, the nearest target
changed from one triangle to the other.
The contact energy remained continuous,
but the horizontal force changed abruptly.
This shows that choosing the nearest target triangle can
introduce force discontinuities even when the quartic
contact law itself is smooth.

\subsubsection{Contact patch test}

\begin{figure}[H]
\centering
\includegraphics[width=0.8\linewidth]{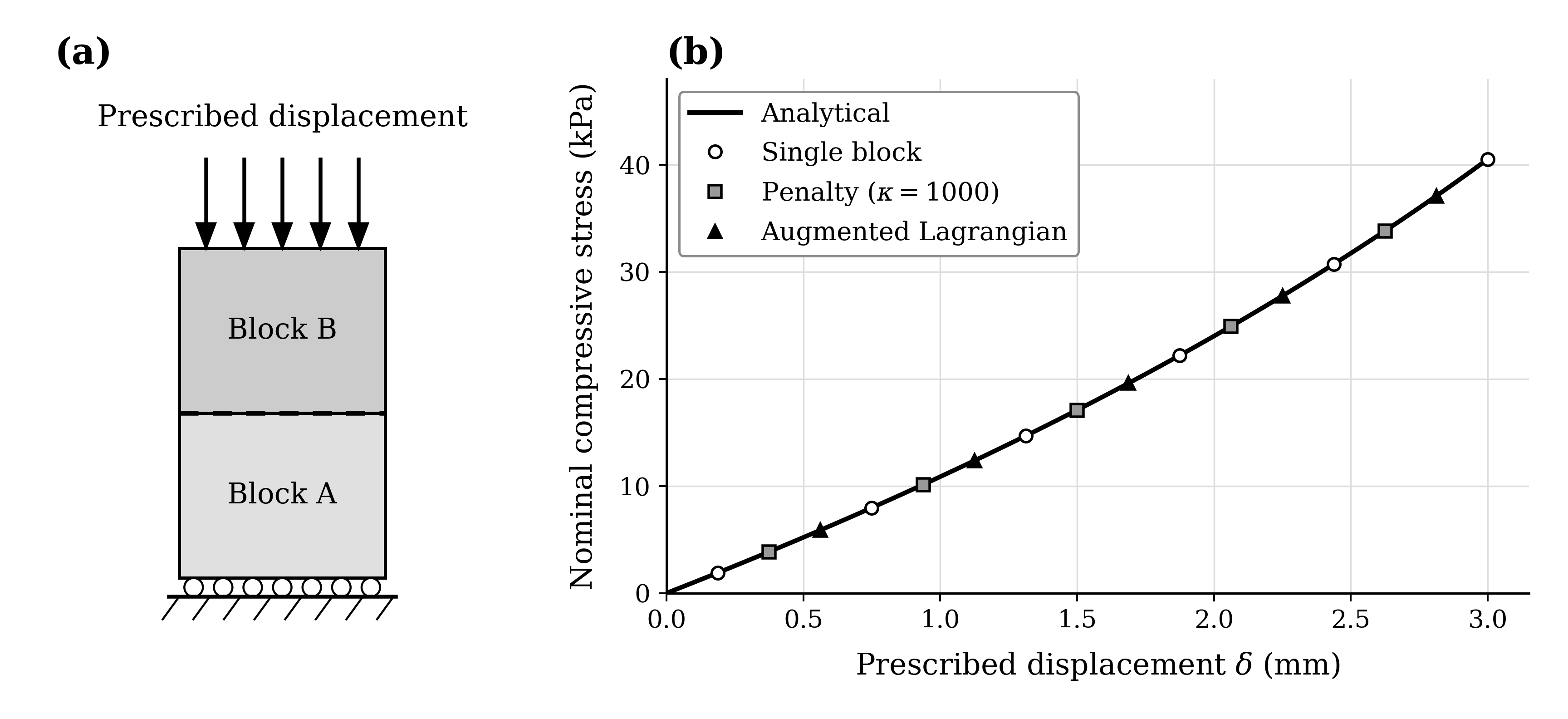}
\caption{Contact patch test under displacement-controlled compression:
(a)~geometry and boundary conditions; and
(b)~nominal compressive stress versus
prescribed downward displacement
for the analytical, single-block, quadratic-penalty
and augmented-Lagrangian solutions.}
\label{fig:patch}
\end{figure}

Uniform force transmission was examined using two
$\SI{10}{\milli\metre}\times\SI{10}{\milli\metre}
\times\SI{5}{\milli\metre}$ neo-Hookean blocks
placed one above the other to form
a \SI{10}{\milli\metre} cube (Fig.~\ref{fig:patch}a).
The material parameters were $E=\SI{100}{\kilo\pascal}$
and $\nu=0.30$.
A quarter-symmetry tetrahedral mesh was used.
Compression was imposed by incrementally
prescribing a downward displacement of the upper block's
top surface, reaching \SI{3}{\milli\metre}
($30\%$ nominal compression).
The bottom surface of the lower block
was constrained in the vertical direction
and free to slide laterally.
Symmetry conditions were applied to the cut faces,
and exterior lateral surfaces were traction-free.
The contact surfaces were frictionless and independently meshed.

Three configurations were compared: a single block without
a contact interface, two blocks with quadratic-penalty contact
($\kappa=1000$), and two blocks with augmented-Lagrangian
contact using the same penalty and $12$ augmentation passes.
The single-block model provided the numerical reference,
while the analytical solution was a uniform uniaxial-stress state. Among the three configurations, the
quadratic-penalty solution had the largest force error.
It differed from the analytical compressive force by at most
$0.09\%$ over the prescribed displacement range
(Fig.~\ref{fig:patch}b). At the final compression, the single-block and augmented-Lagrangian
solutions matched the analytical strain $E_{33}=-0.2550$ to the reported
precision; the quadratic-penalty relative strain error was approximately
$0.06\%$.
In all three simulations, the element-wise axial strain was nearly uniform,
with a standard deviation of order $10^{-8}$. Thus, the axial strain was nearly constant throughout
the blocks, consistent with the uniform analytical
deformation.

\subsubsection{Self-contact of a pressurized ring}
\label{sec:self}

Self-contact was assessed on a circular ring using the same shell
element and quartic point-to-triangle contact formulation as
in the valve simulations.
The ring had radius $R=\SI{10}{\milli\metre}$ and thickness
$t=\SI{1.2}{\milli\metre}$.
Its isotropic strain-energy density was
\[
\psi(\bm E)
=
\frac{\lambda}{2}\bigl(\operatorname{tr}\bm E\bigr)^2
+\mu\,\bm E:\bm E,
\]
where $\bm E$ is the Green--Lagrange strain tensor.
Young's modulus $E=\SI{100}{\kilo\pascal}$ and
Poisson's ratio $\nu=0$ give $\lambda=0$ and
$\mu=\SI{50}{\kilo\pascal}$.
Setting $\nu=0$ removes Poisson coupling
to facilitate comparison with the two-dimensional reference. The shell was discretized using $120$ triangular elements.
A $0.5\%$ radial imperfection made the initial ring slightly
oval and guided the buckling pattern. The out-of-plane displacement was constrained by $u_z=0$,
while in-plane motion remained unconstrained. Equilibrium solutions for this benchmark were obtained by direct
energy minimization up to $p=\SI{144}{\pascal}$.

An independent two-dimensional reference used $120$ straight
segments to approximate the continuum theory of an
inextensible ring under external pressure
\citep{tadjbakhsh1967ring,djondjorov2011ring}.
The reference was formulated per unit
out-of-plane width, with $I=t^3/12$.
Its continuum bending and pressure contributions
to the potential were
\begin{equation}
\Pi_{\mathrm e}
=
\frac{EI}{2}\int_0^L(\kappa-\kappa_0)^2\,\mathrm ds+pA,
\qquad L=2\pi R,
\label{eq:ring-elastica}
\end{equation}
where $\kappa$ and $\kappa_0=1/R$ are the current and initial
centerline curvatures, and $A$ is the enclosed area.
In the segmented model, changes in segment angles determined
the bending energy, a segment-length penalty maintained
near inextensibility, and a quadratic node-pair penalty
activated contact at separations below $t$.
This reference therefore provides a comparison independent
of the shell and quartic contact implementations.

\begin{figure}[H]
\centering
\includegraphics[width=0.96\linewidth]
{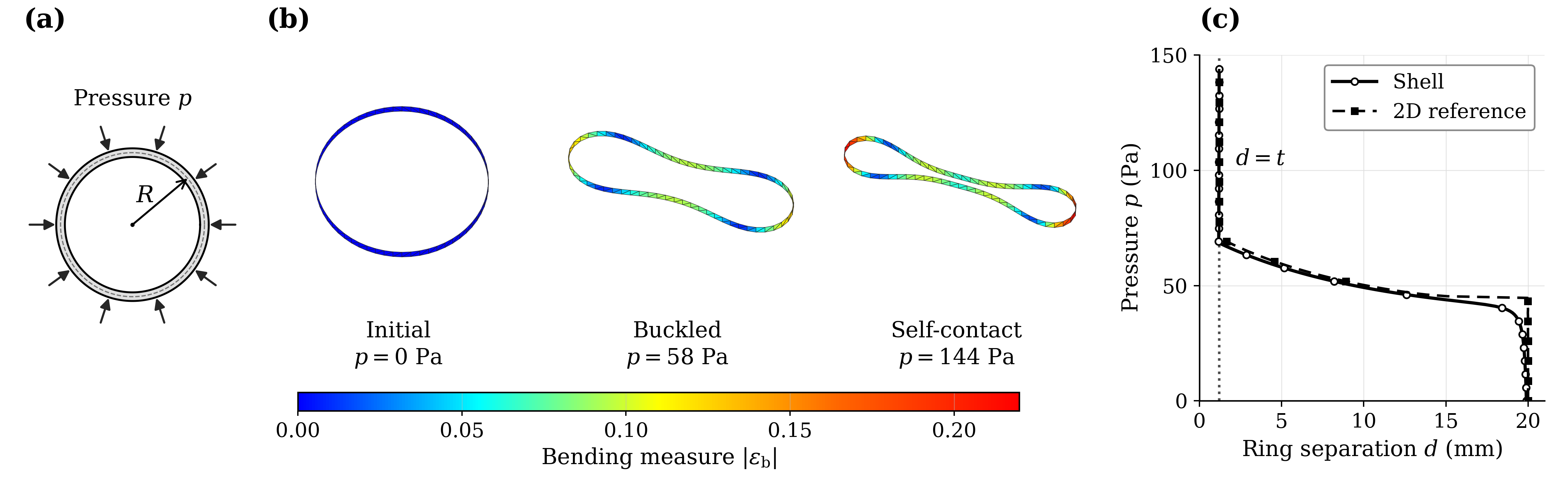}
\caption{Self-contact of a shell ring under uniform external pressure.
(a)~Geometry and boundary conditions;
(b)~initial, buckled and self-contact shell
midsurfaces colored by the curvature-based bending measure
$|\varepsilon_{\mathrm b}|=(t/2)|\kappa-\kappa_0|$,
where $\kappa$ and $\kappa_0$ are the current and initial centerline curvatures; and
(c)~pressure versus ring separation $d$
for the shell and the two-dimensional segmented-ring reference.}
\label{fig:self}
\end{figure}

The nearly circular shell evolved into an oval and then formed
the expected two-lobe self-contact configuration
\citep{flaherty1972ring,djondjorov2011ring}.
The ring separation $d$ was defined as the
minimum distance between initially opposite points on the
ring centerline, as plotted in Fig.~\ref{fig:self}c.
At the final load, $d=\SI{1.20}{\milli\metre}$,
approximately $0.37\%$ below the reference separation
$t=\SI{1.2}{\milli\metre}$.
An independent triangle-intersection check detected no
intersections between non-adjacent midsurface triangles.
The shell and two-dimensional reference differed by
approximately $0.05\%$ in ovalization,
$(r_{\max}-r_{\min})/R$, and
$0.09\%$ in ring separation $d$.
Their final circumferences were respectively
$0.88\%$ and $0.02\%$ below $2\pi R$, reflecting the
greater circumferential contraction permitted by the shell.

\subsection{Moment-loaded roll-up: large-rotation shell bending}
\label{sec:rollup}

The three-node MITC3 seven-parameter extensible-director shell
was evaluated under large rotation using the moment-loaded
roll-up benchmark
\citep{sze2004benchmark,simovuquoc1986beam}.
A clamped strip of length $L=\SI{12}{\milli\metre}$,
width $b=\SI{1}{\milli\metre}$ and thickness
$t=\SI{0.1}{\milli\metre}$ was loaded by an end moment.
The material was St Venant--Kirchhoff,
with $E=\SI{1.2}{\giga\pascal}$ and $\nu=0$.
Motion was restricted to the $x$--$z$ plane.
Setting $\nu=0$ removes Poisson-induced transverse curvature,
permitting comparison with analytical pure bending.
With $I=bt^3/12$, the bending stiffness was
$EI=\SI{0.1}{\newton\square\milli\metre}$. The thickness stretch was fixed at the clamped edge
and allowed to vary at all other nodes.
The EAS mode was disabled.
This benchmark therefore assesses the bending response
without EAS relaxation; the formulation with EAS is assessed
using plate inflation (Section~\ref{sec:plate}).

The analytical solution is a circular arc with uniform
curvature $\kappa=M/EI$.
Defining the total rotation as $\phi=ML/EI$,
the tip coordinates are
\[
\frac{x_{\mathrm{tip}}}{L}=\frac{\sin\phi}{\phi},
\qquad
\frac{z_{\mathrm{tip}}}{L}=\frac{1-\cos\phi}{\phi}.
\]
The strip forms a complete circle at
$M=2\pi EI/L=\SI{52.4}{\milli\newton\milli\metre}$.

\begin{figure}[H]
\centering
\includegraphics[width=0.96\linewidth]
{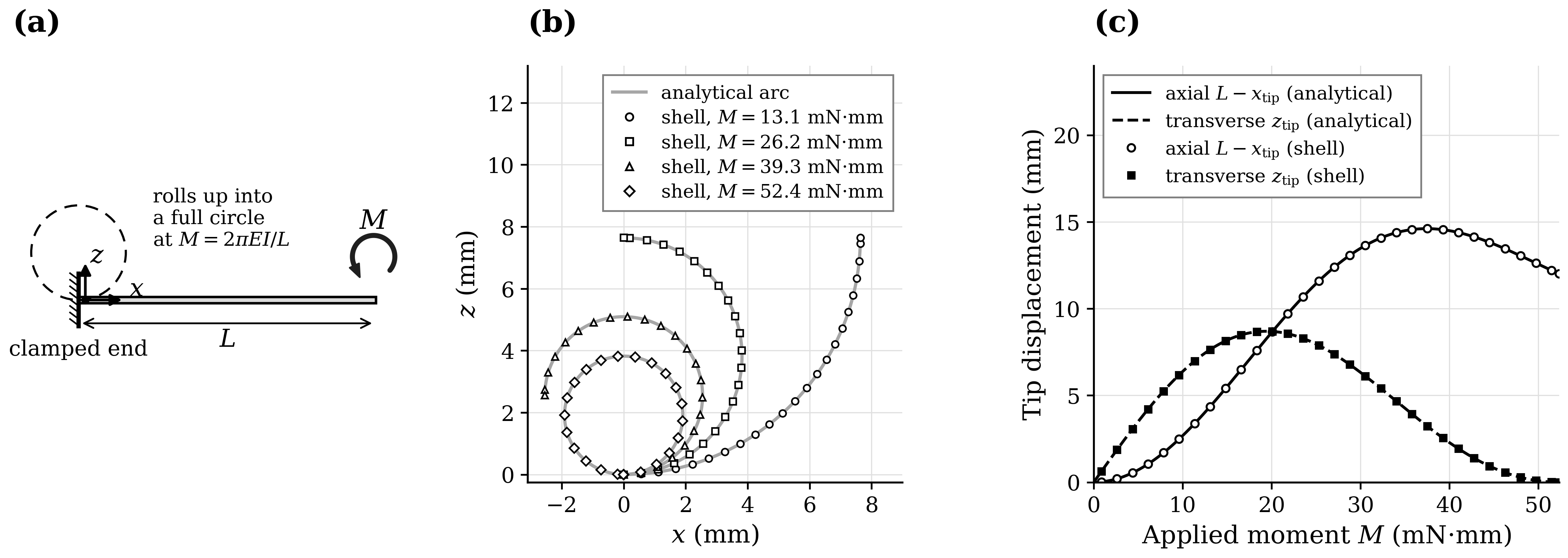}
\caption{Moment-loaded roll-up.
(a)~Geometry and boundary conditions;
(b)~deformed shell midline at four moment levels compared
with the analytical circular arcs; and
(c)~axial $L-x_{\mathrm{tip}}$
and transverse $z_{\mathrm{tip}}$
tip displacements versus applied moment.}
\label{fig:rollup}
\end{figure}

In the small-moment limit, the shell reproduced the
analytical cantilever deflection
$z_{\mathrm{tip}}=ML^2/(2EI)$
with a relative error of $3.3\times10^{-4}\%$. For $32$, $64$ and $128$ triangular elements, the maximum
normalized tip errors were $9.03\%$, $0.99\%$ and
$0.05\%$, respectively, compared with the analytical solution.
In each case, the maximum occurred at complete roll-up.
With $128$ elements, the mean error over the loading ramp
was $0.02\%$, and the final closure gap between the tip and the clamped origin was
\SI{6.56e-3}{\milli\metre}. The tip-displacement curves also followed the analytical
response (Fig.~\ref{fig:rollup}c).
The transverse displacement reached approximately
\SI{8.7}{\milli\metre} before returning close to zero. During roll-up, the tip temporarily moved past the clamped
origin in the $x$ direction, giving $x_{\mathrm{tip}}<0$.
It then returned near the origin as the circle closed, reproducing the analytical reversal of
tip motion under large rotation.

\subsection{Circular-plate inflation:
shell response and mesh sensitivity}
\label{sec:plate}

Circular-plate inflation examined the combined
membrane and bending response under large deformation.
A circular plate with
$R=\SI{7.5}{\milli\metre}$ and
$t=\SI{0.5}{\milli\metre}$
was inflated by uniform pressure to
\SI{35}{\kilo\pascal}.
At every boundary node, displacement was
constrained in all three directions, while director rotations,
nodal thickness stretch, and element-local enhanced-strain
variables remained free.
The material used an isochoric
Mooney--Rivlin model with
$C_1=\SI{80}{\kilo\pascal}$,
$C_2=\SI{20}{\kilo\pascal}$ and a volumetric penalty
$\tfrac12 K(\ln J)^2$, where $K/\mu=50$ and
$\mu=2(C_1+C_2)$. Five uniformly refined meshes contained $19$, $61$, $217$,
$817$ and $3169$ nodes, with the mean element size
reduced to approximately half its previous
value at each refinement level.
The final apex displacement increased monotonically from
$7.45$ to \SI{9.69}{\milli\metre}.
The finest-mesh value was $0.92\%$ higher than the preceding
mesh level.
On the fourth mesh level, increasing the pressure ramp from
$24$ to $32$ steps changed the final apex displacement by
\SI{0.0022}{\milli\metre} ($0.02\%$).

\begin{figure}[H]
\centering
\includegraphics[width=0.96\linewidth]
{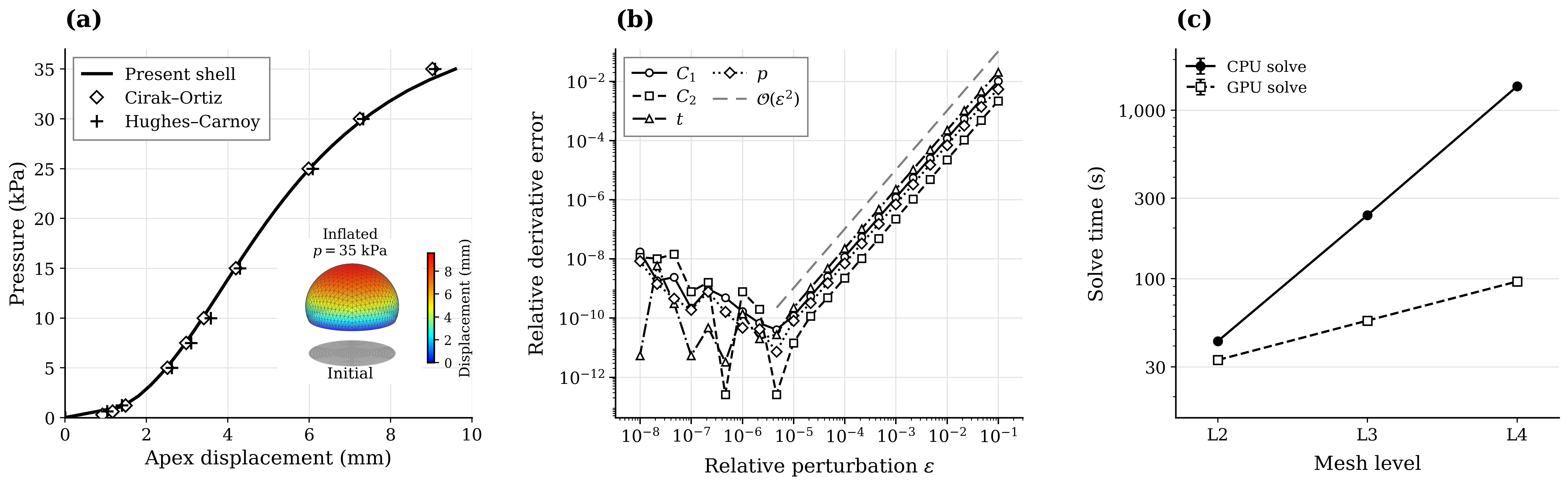}
\caption{Circular-plate inflation and derivative verification.
(a)~Pressure--apex displacement response on the fourth mesh
($817$ nodes; $32$ steps), compared with Cirak--Ortiz and
Hughes--Carnoy results with initial and inflated configurations
at \SI{35}{\kilo\pascal}.
(b)~Relative errors between implicit and central finite-difference
apex-displacement derivatives for $C_1$, $C_2$, $t$, and $p$
versus perturbation size $\varepsilon$ at \SI{20}{\kilo\pascal}.
The gray dashed line indicates second-order convergence. (c)~CPU--GPU solve times for circular-plate
inflation using 32 pressure steps to \SI{35}{\kilo\pascal}.
Points show medians of five runs. The time axis is logarithmic.
CPU: 12 Xeon Gold 5317 cores; GPU: one NVIDIA A100-SXM4-40GB.}
\label{fig:plate_path}
\end{figure}

Figure~\ref{fig:plate_path}a compares the fourth-level
$32$-step response with the
Cirak--Ortiz subdivision-shell and Hughes--Carnoy
axisymmetric-solid results, both
obtained from
Fig.~8 of \citet{cirak2001subdivision}
\citep{hughes1983shell}.
At \SI{35}{\kilo\pascal}, the computed apex displacement
was\SI{9.59}{\milli\metre}, compared with approximately
\SI{9.03}{\milli\metre} and \SI{9.10}{\milli\metre},
respectively:
$6.3\%$ and $5.4\%$ higher. The published models used different
formulations, so these differences
cannot be attributed to discretization error alone.
Incompressibility was enforced directly by Cirak and
Ortiz and through a penalty by Hughes and Carnoy.
On the fourth-level mesh, increasing
$K/\mu$ from $50$ to $1000$ reduced the apex displacement
by $11.9\%$.

\subsubsection{Differentiability verification}
\label{sec:diff}

Differentiability of the converged response was verified using
the circular plate at $p=\SI{20}{\kilo\pascal}$.
On the second mesh level, the relative residual was
$3.2\times10^{-13}$, and the tangent for the unconstrained
degrees of freedom was positive definite.
Derivatives of the converged apex displacement
$W$ with respect to $C_1$, $C_2$, thickness $t$, and pressure
$p$ were obtained by differentiating the equilibrium equations.
These implicit derivatives,
$\mathrm dW/\mathrm dC_1$,
$\mathrm dW/\mathrm dC_2$,
$\mathrm dW/\mathrm dt$, and
$\mathrm dW/\mathrm dp$,
were compared with central finite differences from
independent nonlinear re-solves.
For the finite-difference calculations,
each parameter was varied separately using $22$ relative
perturbation sizes between $10^{-1}$ and $10^{-8}$.
The bulk modulus and drilling-penalty coefficient were held
fixed during these perturbations.
For a differentiable equilibrium branch with nonsingular
\(\bm R_{\bm y}\), implicit differentiation of the complete equations
\(\bm R(\bm y,q)=\bm0\) gives
\[
\frac{\mathrm d\bm y}{\mathrm dq}
=-\bm R_{\bm y}^{-1}\bm R_q.
\]
The Jacobian in this relation is that of the equilibrium equations,
including internal-field coupling and the geometry dependence of the
pressure force. It is distinct from a regularized correction operator
used to reach equilibrium.

As the perturbation size decreased, the relative errors
between implicit and central finite-difference apex-displacement derivatives
initially decreased approximately quadratically,
as expected for central finite differences.
At very small perturbations, the finite-difference calculation
subtracts nearly equal displacements and divides by a very
small parameter increment.
This amplifies round-off and nonlinear-solve errors,
causing the derivative errors to increase
(Fig.~\ref{fig:plate_path}b). For each parameter, the derivative of the
full equilibrium solution agreed with finite differences,
with a minimum relative error of $3.4\times10^{-10}$ or smaller over the tested perturbation sizes.
Similar agreement was obtained on the next mesh level.
These results verify implicit solution derivatives for the
tested contact-free shell equilibrium solutions.
The quartic contact-energy derivatives were tested separately
in Section~\ref{sec:traction}.

\subsubsection{Computational performance}
\label{sec:perf}

The circular-plate benchmark was used to compare CPU and GPU
execution of the same matrix-free solver using a 32-step
pressure ramp to \SI{35}{\kilo\pascal}.
Levels L2--L4 contained $451$, $1\,759$, and $6\,967$
free degrees of freedom, respectively, including nodal
variables and one enhanced-strain variable per element.
The solver evaluated Hessian--vector products using AD
and Jacobi-preconditioned CG.

CPU and GPU runs were performed on the same compute node
using 12 physical Intel Xeon Gold 5317 cores and one
NVIDIA A100-SXM4-40GB, respectively.
All runs used float64 arithmetic with Python~3.12.3,
JAX/jaxlib~0.10.2, NumPy~2.4.2, and SciPy~1.17.0.
Five independent timed runs were performed for each device
and mesh level, giving 30 timed runs.
Solve times include first-use JIT compilation and nonlinear
and linear solution, but exclude imports and mesh generation. Convergence was checked using the implemented residual indicator,
\[
\rho=
\frac{\lVert\bm g\rVert}
{\max(1,\lVert\bm f\rVert)},
\]
where $\bm g$ is the numerical energy-gradient vector on the
unknown variables and $\bm f$ is the numerical pressure-load
vector evaluated at the start of each load step.
Because the unknowns include translations, rotations, and
internal variables, this indicator uses the benchmark's
fixed mm--kPa coordinate convention.
All 30 timed runs satisfied $\rho\le10^{-6}$ at every load
step and produced final apex displacements agreeing to the
reported precision: $8.46$, $9.32$, and
\SI{9.59}{\milli\metre} for L2--L4, respectively.

Median CPU solve times were $42.59$, $238.16$, and
\SI{1383.00}{\second}, compared with $32.94$, $56.42$, and
\SI{96.25}{\second} on the GPU for L2--L4, respectively
(Fig.~\ref{fig:plate_path}c).
The corresponding ratios of CPU to GPU median times were
$1.29$, $4.22$, and $14.37$.
These speedups are relative to the specified 12-core CPU
baseline on the tested hardware.
Recorded Krylov iteration counts were identical across CPU
repeats but varied by approximately $10$--$13\%$ across GPU
repeats. The timing ratios therefore reflect differences in numerical
work as well as execution speed.

\subsection{Application: patient-specific aortic valve simulations}
\label{sub:valve}

The formulation was applied to three patient-specific aortic valves:
two tricuspid and one bicuspid. All cases used the same Lee--Sacks
parameters and solver, with patient-specific geometry and annulus motion.

\subsubsection{Geometry, discretization and boundary conditions}

Leaflet geometries were reconstructed from manual segmentations
in the open reference configuration, following our previous workflow
\citep{nakhaei2026registration}. The open reference configuration
was assumed to be stress-free. The collection and analysis of the clinical imaging data were approved by the Institutional Review Boards at the
Children's Hospital of Philadelphia and the University of
Pennsylvania. Medial surfaces were extracted using an interactive skeletonization
tool in 3D Slicer \citep{fedorov20123d,herz2024open}, smoothed while
preserving the attachment boundaries and free edges, and discretized
with triangular shell elements. ACVD remeshing was used where needed
\citep{valette2008acvd}.
All leaflets had a uniform thickness of $\SI{1.2}{\milli\metre}$.
The nominal thickness follows the previous modeling workflow
\citep{nakhaei2026registration}, informed by the autopsy study of
\citet{sahasakul1988thickness}. It was not measured separately for these patients.
A uniform pressure load was applied to the ventricular face and
ramped to $\SI{10}{\kilo\pascal}$, approximately $\SI{75}{mmHg}$.
Prescribed annulus displacements were obtained from image registration
\citep{nakhaei2026registration}. The mean annulus displacement was
subtracted from each annulus nodal displacement to remove overall
translation while retaining relative annulus motion.
These displacements were ramped with the pressure.

\subsubsection{Leaflet strain, stress and computational performance}

\begin{figure}[H]
\centering
\includegraphics[width=0.7\linewidth]
{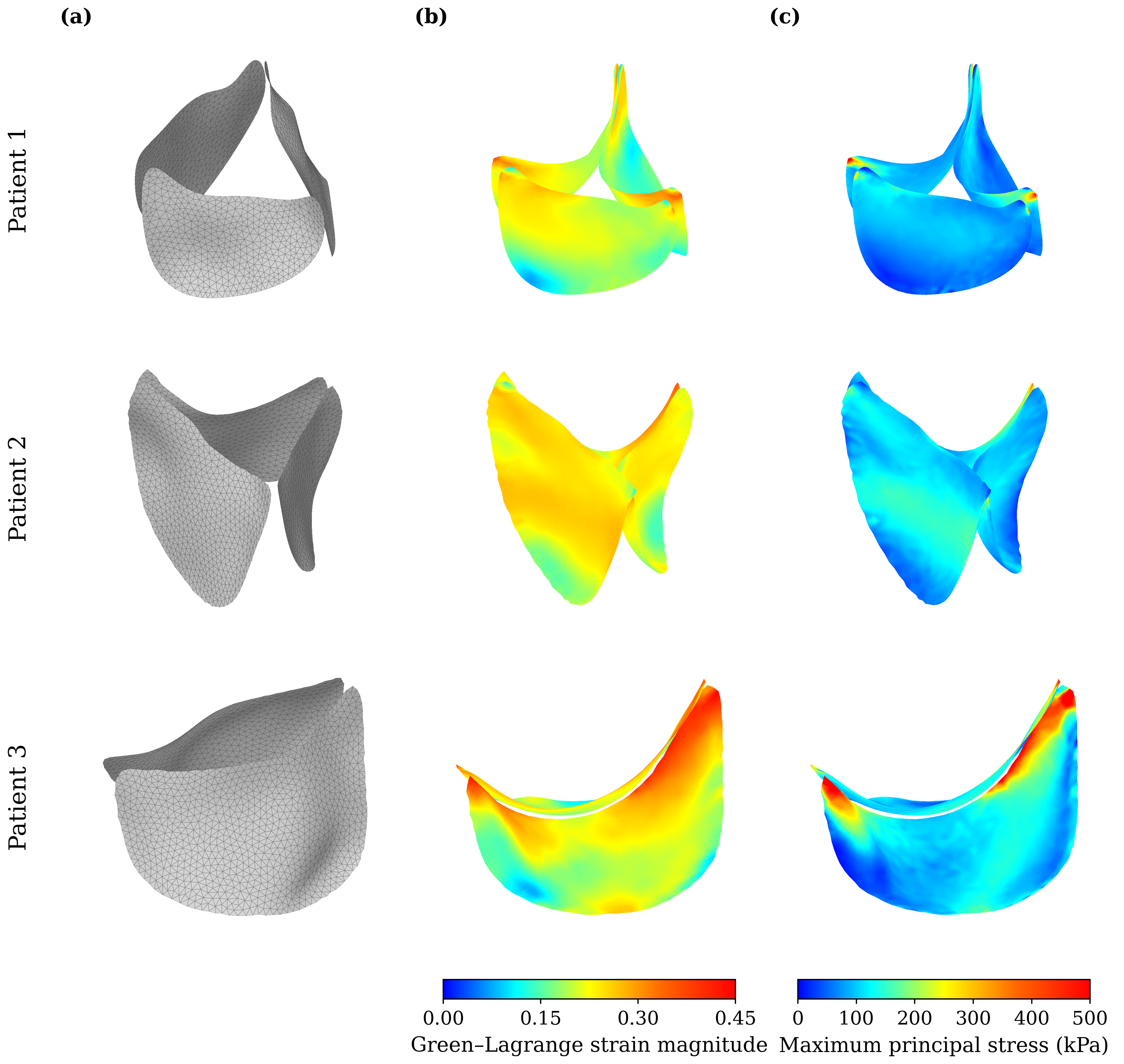}
\caption{Patient-specific aortic-valve simulations.
Rows show patient~1 (tricuspid), patient~2 (tricuspid),
and patient~3 (bicuspid).
Columns show (a)~the open reference configuration,
(b)~Green--Lagrange strain magnitude and
(c)~estimated maximum principal Cauchy
stress on the closed configurations at
$p=\SI{10}{\kilo\pascal}$.}
\label{fig:patient_valves}
\end{figure}

All three CPU simulations reached \SI{10}{\kilo\pascal}.
Patients~1 and 2 completed $40$ load steps
without rejection; patient~3 completed $45$ accepted steps
after one rejected attempt.
The closed configurations are shown in
Fig.~\ref{fig:patient_valves}.
After final internal-field refinement and recalculation
of the pressure force from the deformed geometry,
the relative equilibrium residuals
$\lVert\bm r\rVert/\lVert\bm F_{\mathrm p}\rVert$
were below $10^{-4}$ in all three cases, satisfying
the convergence criterion of $10^{-3}$.

The median Green--Lagrange strain magnitudes  ($\lVert\bm E\rVert$) were $0.19$,
$0.23$, and $0.21$ for patients~1--3, respectively.
Higher strains occurred near the commissures and the
free edges involved in coaptation.
The maximum strain magnitudes were
$0.373$, $0.362$, and $0.450$, respectively. Maximum principal Cauchy stress ($\sigma_1$) was estimated by evaluating
the constitutive response at each element's averaged
Green--Lagrange strain tensor. Median stresses were \SI{81.0}{\kilo\pascal},
\SI{101.2}{\kilo\pascal}, and \SI{108.8}{\kilo\pascal}
for patients~1--3, respectively.
The maximum stress were \SI{222.9}{\kilo\pascal},
\SI{191.4}{\kilo\pascal}, and \SI{494.5}{\kilo\pascal}. GPU execution reduced the median runtime by
factors of $5.79$, $5.27$, and $6.52$ for patients~1--3,
respectively, on the tested hardware.
Table~\ref{tab:valve} summarizes the mechanical results
and CPU--GPU performance, while
Fig.~\ref{fig:patient_valves} shows the reference and
closed configurations with the strain and stress fields.

\begin{table}[H]
\centering
\caption{Patient-specific aortic-valve closure at
\SI{10}{\kilo\pascal}.
Peak RSS is the maximum host-process resident memory
over accepted timed runs.
CPU and GPU runs were performed sequentially on one node
using $12$ physical Intel Xeon Gold 5317 cores and one
NVIDIA A100-SXM4-40GB.}
\label{tab:valve}
\small
\begin{tabularx}{\linewidth}{@{}Xccc@{}}
\toprule
 & patient~1 & patient~2 & patient~3 \\
 & (tricuspid) & (tricuspid) & (bicuspid) \\
\midrule
number of leaflets
    & 3 & 3 & 2 \\
global dof
    & $11\,256$ & $11\,207$ & $11\,200$ \\
$(c_0,c_1,K)$ (\si{\kilo\pascal})
    & \multicolumn{3}{c}{$(67,13,5000)$} \\
$c_2$ (dimensionless)
    & \multicolumn{3}{c}{$38$} \\
\midrule
final relative residual
    & $8.31\times10^{-5}$
    & $6.95\times10^{-5}$
    & $6.19\times10^{-5}$ \\
\midrule
median $\lVert\bm E\rVert$
    & $0.19$ & $0.23$ & $0.21$ \\
maximum $\lVert\bm E\rVert$ & $0.37$
    & $0.36$
    & $0.45$ \\
median $\sigma_1$ (\si{\kilo\pascal})
    & $81.00$
    & $101.20$
    & $108.80$ \\
maximum $\sigma_1$ (\si{\kilo\pascal})
    & $222.90$
    & $191.40$
    & $494.50$ \\
maximum displacement (\si{\milli\metre})
    & $8.44$ & $9.53$ & $14.74$ \\
\midrule
CPU wall time, median (\si{\minute})
    & $87.88$ & $77.75$ & $165.65$ \\
GPU wall time, median (\si{\minute})
    & $15.17$ & $14.77$ & $25.40$ \\
CPU/GPU speed-up
    & $5.79$ & $5.27$ & $6.52$ \\
peak host RSS, CPU/GPU (\si{\gibi\byte})
    & $3.35/4.37$ & $3.26/4.22$ & $3.41/4.12$ \\
\bottomrule
\end{tabularx}
\end{table}

\subsubsection{Mesh-sensitivity study}
\label{sec:meshconv}

\begin{figure}[H]
\centering
\includegraphics[width=\linewidth]{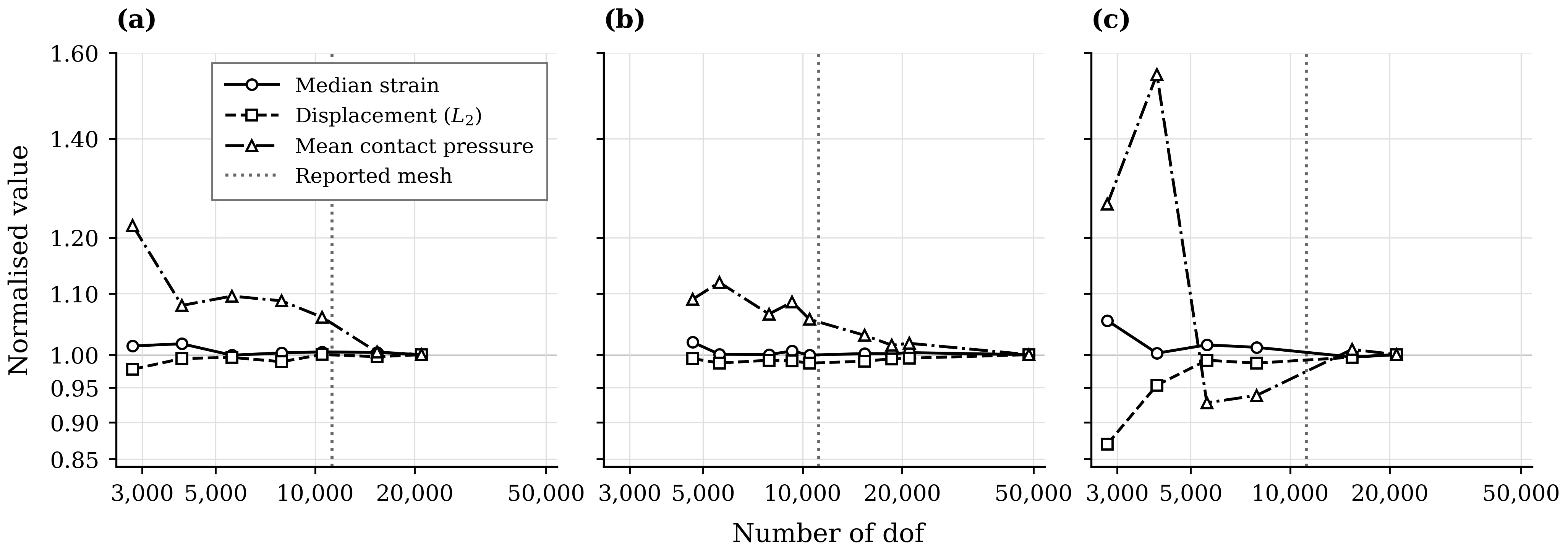}
\caption{Mesh sensitivity for (a)~patient~1,
(b)~patient~2, and (c)~patient~3.
Each quantity is normalized by its value on the patient's
finest accepted mesh. Displacement ($L_2$) denotes root-mean-square nodal displacement
from the open reference to the closed configuration.
Dotted lines mark the dof counts used for the reported simulations.}
\label{fig:meshconv}
\end{figure}

For each patient, closure was repeated on independently generated
ACVD meshes \citep{valette2008acvd}.
Of $24$ simulations, $22$ reached
\SI{10}{\kilo\pascal}, completed refinement without an
unrecovered stall, and satisfied
$\lVert\bm r\rVert/\lVert\bm F_{\mathrm p}\rVert\le10^{-3}$.
The patient~2 mesh with $2\,800$ dof stopped at
\SI{2.99}{\kilo\pascal}.
The patient~3 mesh with $10\,500$ dof reached the target
pressure but was excluded because its final relative
residual was $4.08\times10^{-3}$. Three quantities were examined: median Green--Lagrange
strain magnitude, root-mean-square nodal displacement,
and reference-area-weighted mean contact
pressure over the active contact region
(Fig.~\ref{fig:meshconv}).
Maximum strain was excluded because its discrete value
is sensitive to local refinement near the constrained commissures.

For each patient, we selected the nearest accepted meshes
with fewer and more dof than the mesh reported in
Table~\ref{tab:valve}.
Each quantity on these meshes was compared
with its value on the patient's finest accepted mesh.
For each quantity, the larger absolute relative difference
from the two neighboring meshes was reported as a percentage
of the finest-mesh value. The differences for patients~1--3 were $0.43\%$, $0.17\%$,
and $1.15\%$ for median strain; $0.32\%$, $1.32\%$,
and $1.31\%$ for root-mean-square nodal displacement;
and $6.00\%$, $5.65\%$, and $6.16\%$ for mean contact pressure. Median strain and displacement therefore showed limited
mesh sensitivity near the selected resolutions,
while contact pressure was more sensitive.
Because the meshes were generated independently, changes
with increasing dof were not necessarily monotonic.

\subsubsection{Numerical-sensitivity checks}
\label{sec:verification-audit}

Additional simulations examined sensitivity to the
pressure-force update tolerance and through-thickness
quadrature. At full load, pressure-force updates
stopped when,
\[
\frac{
\lVert\bm F_{\mathrm p}^{(j)}
-\bm F_{\mathrm p}^{(j-1)}\rVert
}{
\lVert\bm F_{\mathrm p}^{(j)}\rVert
}
\le\varepsilon_{\mathrm p},
\]
where $j$ is the update number and
$\varepsilon_{\mathrm p}$ is the pressure-force tolerance. For patient~2, the simulation was repeated with
$\varepsilon_{\mathrm p}=10^{-5}$, $10^{-4}$, and $10^{-3}$.
Relative to the $10^{-4}$ case, the monitored
strain, displacement, and contact-distance measures
changed by at most $0.008\%$.
The reported simulations used $\varepsilon_{\mathrm p}=10^{-4}$,
which required $10$ pressure-force updates
in this test, compared with $14$ for $10^{-5}$.
The looser tolerance of $10^{-3}$ produced a relative
equilibrium residual of $8.90\times10^{-4}$, close to
the reporting threshold of $10^{-3}$. In all three patient models, the standard two-point
Gauss rule through the thickness was compared with
four-point Gauss and five-point Lobatto rules,
retaining three in-plane integration locations.
The Lobatto rule also samples the shell faces.
Relative to the two-point rule,
the largest changes across the three patients were
$0.82\%$ in median Green--Lagrange strain magnitude, $0.59\%$ in nodal
displacement magnitude, and $0.10\%$ in the
minimum distance from a source-triangle
centroid to an opposing triangle. These checks showed limited sensitivity of the monitored mechanical quantities to the tested pressure-force tolerances and quadrature rules.


\section{Discussion}\label{sec:discussion}

We developed shell and contact implementations
for nonlinear valve mechanics using computational building blocks from
tatva's energy-based AD framework \citep{pundir2026versatile}.
We compared constitutive and shell responses with reference solutions
\citep{sze2004benchmark,simovuquoc1986beam,cirak2001subdivision,hughes1983shell}
and checked contact-force calculations in dedicated benchmarks.
Contact-free sensitivity analyses further showed agreement between
derivatives of converged shell solutions and finite-difference estimates.
The implementation supported nonlinear forward simulation of three
patient-specific aortic valves with tricuspid and bicuspid geometries,
and the matrix-free solver achieved shorter runtimes on the GPU
than on the CPU.

\subsection{Patient-specific valve simulations and numerical sensitivity}

All three baseline valve simulations reached \SI{10}{\kilo\pascal},
with final relative residuals below \(10^{-4}\) for the original,
unregularized equilibrium equations. Median Green--Lagrange strain
magnitudes were approximately \(19\%\), \(23\%\), and \(21\%\) for
patients~1--3, respectively (Table~\ref{tab:valve}), with higher
magnitudes near the commissures and contacting free edges.
This spatial pattern was qualitatively consistent with observations
from our previous registration study \citep{nakhaei2026registration}.
That study reported leaflet-type cohort means of
\(17.3\%\)--\(17.7\%\) for tricuspid valves and
\(11.9\%\)--\(16.5\%\) for bicuspid valves
\citep{nakhaei2026registration}. Jermihov et al.\ reported maximum in-plane logarithmic strains
of \(15.7\%\) for the tricuspid valve and
\(18.1\%\)--\(25.5\%\) across bicuspid geometries using their
baseline material parameters \citep{jermihov2011effect}.
Differences in strain estimates, patient populations,
geometry, and loading limit direct quantitative comparison among
these studies.

Mean contact pressure was more sensitive to mesh resolution
than median strain and root-mean-square displacement
(Section~\ref{sec:meshconv}), suggesting that mesh selection
should account for the mechanical quantity of interest.
For the accepted meshes neighboring the reported resolutions,
the largest relative differences from each patient's finest
accepted mesh were \(1.15\%\) in median strain,
\(1.32\%\) in root-mean-square displacement, and
\(6.16\%\) in mean contact pressure.
Because the meshes were generated independently, these
differences reflect changes in both resolution and mesh
distribution.
Additional checks showed that the strain,
displacement, and contact-distance measures varied only slightly
with the tested pressure-force update tolerances and
through-thickness quadrature rules
(Section~\ref{sec:verification-audit}).
For patient~2, the selected \(10^{-4}\) pressure-force tolerance
provided a balance between computational effort and equilibrium
accuracy. Tightening it to \(10^{-5}\) increased the number of
updates, while loosening it to \(10^{-3}\) increased the final
equilibrium residual.

\subsection{Differentiability and computational performance}

For the contact-free plate,
implicit derivatives of the converged apex displacement
agreed with finite differences from independent nonlinear
re-solves for all four tested parameters
(Section~\ref{sec:diff}).
The contact tests separately checked contact-energy
derivatives within fixed contact branches
(Section~\ref{sec:traction}).
Table~\ref{tab:frameworks} summarizes how differentiation is
used in related biomechanical studies.
Adjoint methods \citep{finsberg2018biventricular} and PyTorch-FEA
\citep{liang2023pytorchfea} support parameter estimation, and
energy-trained NNFE models provide rapid predictions after training
\citep{meyer2025nnfe}.
A trained network can remain differentiable with respect to its
inputs, but those derivatives describe the learned
approximation and require their own verification.
The present implementation focuses on shell-based forward valve
simulation, combining large-deformation leaflet mechanics and contact
with matrix-free computation on CPUs and GPUs.
Its use of extensible-director shells distinguishes the mechanical
discretization from the solid-element approaches listed in the table.

\begin{table}[htbp]
\centering
\begingroup
\caption{Functionality demonstrated in selected biomechanical
studies. Entries describe the reported applications rather than all
capabilities of the underlying software. ``Not demonstrated'' refers to
the reported examples; ``not reported'' indicates unspecified GPU use.}
\label{tab:frameworks}
\fontsize{8}{9.5}\selectfont
\setlength{\tabcolsep}{3pt}
\renewcommand{\arraystretch}{1.2}
\begin{tabularx}{\linewidth}{@{}
>{\hsize=1.15\hsize\linewidth=\hsize\RaggedRight\arraybackslash}X
>{\hsize=1.1\hsize\linewidth=\hsize\RaggedRight\arraybackslash}X
>{\hsize=1.0\hsize\linewidth=\hsize\RaggedRight\arraybackslash}X
>{\hsize=0.8\hsize\linewidth=\hsize\RaggedRight\arraybackslash}X
>{\hsize=1.0\hsize\linewidth=\hsize\RaggedRight\arraybackslash}X
>{\hsize=1.25\hsize\linewidth=\hsize\RaggedRight\arraybackslash}X
>{\hsize=0.7\hsize\linewidth=\hsize\RaggedRight\arraybackslash}X@{}}
\toprule
Study / application
& Computational approach
& Elements
& Anisotropic hyperelasticity
& Contact scope
& Derivative calculations
& GPU use \\
\midrule

\citet{finsberg2018biventricular}\newline Ventricles
& FEniCS / dolfin-\allowbreak adjoint
& Tetrahedral solids
& Yes
& Not demonstrated
& Adjoint equilibrium gradients for parameter estimation
& CPU \\
\addlinespace

\citet{sacks2022neural}\newline Myocardium; aortic leaflet
& Energy-trained NNFE; JAX for leaflet
& 3D solids; NURBS leaflet
& Yes
& Not demonstrated
& Neural-network training AD
& Not reported \\
\addlinespace

\citet{liang2023pytorchfea}\newline Aorta
& PyTorch-FEA / PyTorch
& Hexahedral solids
& Yes
& Not demonstrated
& Stress, stiffness, and inverse-loss AD
& Yes (hybrid) \\
\addlinespace

\citet{meyer2025nnfe}\newline Trileaflet valve
& Energy-trained NNFE / JAX
& 3D NURBS solids
& Yes
& Large-deformation multi-body contact;
self-contact not demonstrated
& Neural-network training AD
& GPU-based; hardware unspecified \\
\addlinespace

Present study\newline Aortic valve
& Matrix-free / JAX; tatva FE foundation
& Extensible-\allowbreak director shells
& Cylinder benchmark; isotropic valves
& Large-deformation leaflet contact;
self-contact in ring benchmark
& Residual and tangent AD;
contact-free plate equilibrium sensitivities;
contact-energy derivatives
& Yes (hybrid) \\

\bottomrule
\end{tabularx}
\endgroup
\end{table}

The CPU--GPU comparisons demonstrate acceleration of the
reported matrix-free implementations on the tested hardware.
For circular-plate inflation, the ratio of CPU to GPU
median solve times increased from \(1.29\) at L2 to
\(14.37\) at L4 (Fig.~\ref{fig:plate_path}(c)).
For the patient-specific valves, the ratios of CPU to GPU
median runtimes were \(5.27\)--\(6.52\)
(Table~\ref{tab:valve}). These ratios are relative to the
specified 12-core CPU baseline. The plate solve times include
first-use JIT compilation, and their ratios reflect differences
in Krylov iteration counts as well as execution speed.
Tatva supports hybrid workflows by allowing JAX-based
computations to be combined with external solvers
\citep{pundir2026versatile}.
The valve speedups were achieved with a hybrid workflow that
combines GPU computation with contact searches, some auxiliary
solves, and host-side solver operations on the CPU.

\subsection{Limitations and future work}

Contact and internal shell fields introduce additional computational
requirements.
Tatva provides matrix-free tangent--vector products and
coloring-based sparse tangent reconstruction from the same
AD-derived residual \citep{pundir2026versatile}.
Coloring requires a sparsity pattern covering both shell and
contact couplings. As contact candidates change, the pattern
must be updated or kept broad enough to cover possible new
interactions beyond the shell-mesh connectivity.
Furthermore, during final valve refinement, global elimination of the
nodal thickness variables can make the reduced tangent
more densely coupled than the original mixed system
(Eq.~\eqref{eq:reduced-tangent}).
Our regularized, matrix-free Newton--Krylov solver accounts for
internal-field coupling without explicitly constructing the
complete reduced tangent.
Auxiliary matrices are retained for preconditioning and
internal-field updates.
Coloring remains a viable alternative when the coupling
pattern is represented correctly; its relative efficiency
for the coupled valve problem would require a matched
comparison of complete solves.
JAX execution also involves memory and compilation considerations.
Reverse-mode AD stores intermediate quantities for derivative
evaluation \citep{jaxdocs2026}.
For contact calculations, fixed-size candidate arrays with
unused entries ignored can avoid dimension changes that would
require recompilation. These arrays can still contain changing
contact pairs, whose couplings must be covered by any coloring
pattern. Changing the closest facet within an existing array
does not itself require recompilation, but can introduce
nonsmooth contact forces (Section~\ref{sec:traction}).
Further runtime reductions may be possible by moving selected
CPU operations to the GPU and improving preconditioning
to reduce the number of Krylov iterations
\citep{knoll2004jfnk}.

In the valve simulations, although the geometries and prescribed annulus motion
are patient-specific, the simulations use isotropic
Lee--Sacks parameters found in the literature, uniform thickness, and an open reference
state assumed to be stress-free.
Representing regional thickness variation in the shell formulation
would account for spatial differences in membrane and bending stiffness
that can influence leaflet deformation and coaptation.
Within the current JAX implementation, structurally motivated anisotropic
constitutive models, such as the octahedral fibrous model of
\citet{parvez2026octahedral}, could represent the effects of fiber
organization and pre-stretch in the valve simulations.
Future work can build on the derivative checks to verify
sensitivities of the converged valve solutions with active contact.
Coupling the JAX-based forward solver with inverse
parameter-estimation methods could further improve patient-specific
simulations by estimating constitutive parameters from imaging-derived
leaflet motion.
The physics-informed neural network approach of \citet{wu2025noninvasive}
demonstrates the potential of image-informed material estimation for this
purpose.
Integrating the present implementation with machine-learning surrogates,
such as the physics-conditioned neural operator developed by
\citet{koohy2026physicsconditioned}, could accelerate repeated response
evaluations within these inverse analyses.

\section{Conclusion}\label{sec:conclusions}

This work demonstrates GPU-accelerated, shell-based simulation of
patient-specific valve mechanics with contact within an AD-based framework.
The benchmarks provided verification evidence for the
tested components and selected coupled problems, including
equilibrium sensitivities for contact-free plates.
For the valve models considered, the matrix-free solver achieved
5.3--6.5-fold speedups on the GPU relative to corresponding CPU runs
on the tested hardware.
These reductions in solution time make repeated analyses of how
geometry, constitutive parameters, and loading influence valve function
more practical.
The implementation provides a foundation for future
gradient-based inverse studies and machine-learning coupling.

\section*{CRediT authorship contribution statement}
\textbf{Mohsen Nakhaei:} Writing -- original draft,
Conceptualization, Methodology, Validation, Visualization, Software, 
Investigation, Formal analysis.
\textbf{Alison M. Pouch:} Writing -- review \& editing,
Supervision, Conceptualization, Methodology, Resources, Funding acquisition.
\textbf{Matthew A. Jolley:} Writing -- review \& editing,
Supervision, Conceptualization, Methodology, Resources, Funding acquisition.
\textbf{Wensi Wu:} Writing -- review \& editing,
Supervision, Conceptualization, Methodology, Resources, Funding acquisition.

\section*{Code availability}
The code used in this study will be made
available upon acceptance of the manuscript.

\section*{Declaration of competing interest}
The authors declare that they have no known
competing financial interests or personal relationships
that could have appeared to influence the work reported
in this paper.

\section*{Acknowledgments}
This work was supported by the Cora Topolewski Pediatric
Valve Center at the Children's Hospital of Philadelphia, a Cardiac Center
Innovation Grant, an Additional Ventures Single Ventricle Research Fund
award (1434814), and the National Institutes of Health through NHLBI grants
K25-HL168235 (W.W.), R01-HL153166 (M.A.J.), and
R01-HL163202 (A.M.P.).

\section*{Declaration of generative AI use}
During the preparation of this manuscript, generative AI tools
were used for language refinement and paraphrasing of selected
text segments to improve clarity and readability, and to assist
with debugging selected parts of the implementation.
All AI-assisted outputs were critically reviewed, verified,
and edited by the authors to ensure accuracy, technical
correctness, and consistency with the intended scientific meaning.

\section*{Data availability}
Data supporting this study are available
on reasonable request to the corresponding authors.
Clinical imaging data are not publicly shared to protect
patient confidentiality.

\bibliographystyle{elsarticle-num-names}
\bibliography{reference}

\end{document}